\documentstyle[emulateapj,psfig]{article}

\begin{document}
\title{Model-Independent Multi-Variable Gamma-Ray Burst Luminosity
Indicator and Its Possible Cosmological Implications}
\author{Enwei Liang$^{1,2}$ and Bing Zhang$^1$}
\affil{$^1$Physics Department, University of Nevada, Las Vegas,
NV89154; \\lew@physics.unlv.edu, bzhang@physics.unlv.edu\\
$^2$Department of Physics, Guangxi University, Nanning 530004, China \\
}

\begin{abstract}
Without imposing any theoretical models and assumptions, we present a multi-variable
regression analysis to several observable quantities for a sample of 15 gamma-ray bursts
(GRBs). The observables used in the analysis includes the isotropic gamma-ray energy
($E_{\gamma,{\rm{iso}}}$), the peak energy of the $\nu F_{\nu}$ spectrum in the rest
frame ($E_{\rm {p}}^{'}$), and the rest frame break time of the optical afterglow light
curves($t_{\rm{b}}^{'}$). A strong dependence of $E_{\gamma, {\rm iso}}$ on $E_{\rm
p}^{'}$ and $t_b^{'}$ is derived, which reads $E_{\gamma, {\rm{iso}}}/10^{52} {\rm
ergs}=(0.85\pm0.21)\times (E^{'}_{\rm {p}}/{\rm 100 \
keV})^{1.94\pm0.17}\times(t^{'}_{\rm{b}}/1{\rm day})^{-1.24\pm0.23}$ in a flat Universe
with $\Omega_M=0.28$ and $H_0=71.3$ km s$^{-1}$ Mpc$^{-1}$. We also extend the analysis
to the isotropic afterglow energies in the X-ray and the optical bands, respectively, and
find that they are essentially not correlated with $E_{\rm {p}}^{'}$ and
$t_{\rm{b}}^{'}$. Regarding the $E_{\gamma, {\rm{iso}}}(E_{\rm {p}}^{'}, t_{\rm{b}}^{'}$)
relationship as a luminosity indicator, we explore the possible constraints on the
cosmological parameters using the GRB sample. Since there is no low-redshift GRB sample
to calibrate this relationship, we weigh the probability of using the relationship in
each cosmology to serve as a standard candle by $\chi^2 $ statistics, and then use this
cosmology-weighed standard candle to evaluate cosmological parameters. Our results
indicate $0.05<\Omega_{\rm{M}}<0.50$ at $1\sigma$ level, with the most possible value of
$\Omega_{\rm{M}}$ being $0.28$. The best value of $\Omega_{\Lambda}$ is 0.64, but it is
less constrained. Only a loose limit of $\Omega_{\Lambda}<1.2$ is obtained at $1\sigma$
level. In the case of a flat Universe, the $1\sigma$ constraints are
$0.13<\Omega_{\rm{M}}<0.49$ and $0.50<\Omega_{\Lambda}<0.85$, respectively. The
decelerating factor ($q$) and its cosmological evolution $(dq/dz)$ are also investigated
with an evolutionary form of $q=q_0+zdq/dz$. The best-fit values are $(q_0,dq/dz)=(-1.00,
1.12)$, with $-2.23<q_0<0.26$ and $-0.07<dq/dz<3.48$ at $1\sigma$ level. The inferred
transition redshift between the deceleration and the acceleration phases is
$0.78^{+0.32}_{-0.23}$ ($1\sigma$). Through Monte Carlo simulations, we find that the GRB
sample satisfying our relationship observationally tends to be a soft and bright one, and
that the contraints on the cosmological parameters can be much improved either by the
enlargement of the sample size or by the increase of the observational precision.
Although the sample may not expand significantly in the {\em Swift} era, a significant
increase of the sample is expected in the long term future. Our similations indicate that
with a sample of 50 GRBs satisfying our multi-variable standard candle, one can achieve a
constraint to the cosmological parameters comparable to that derived from 157 Type Ia
supernovae. Furthermore, the detections of a few high redshift GRBs satisfying the
correlation could greatly tighten the constraints. Identifying high-$z$ GRBs and
measuring their $E'_p$ and $t'_b$ are therefore essential for the GRB cosmology in the
{\em Swift} era.

\end{abstract}


\section {Introduction}
Long gamma-ray bursts (GRBs) are originated from
cosmological distances (Metzger et al. 1997). Their births follow
the star formation history of the universe (e.g. Totani 1997;
Paczynski 1998; Bromm \& Loeb 2002; Lin et al. 2004). GRBs therefore
promise to serve as a new
probe of cosmology and galaxy evolution (e.g., Djorgovski et al.
2003). It is well known that Type Ia supernovae (SNe Ia) are a perfect
standard candle to measure the local universe up to a redshift
of $\sim 2$ (e.g., Riees et al. 2004). Gamma-ray photons (with energy
from tens of keV to several MeV) from GRBs are almost immune to dust
extinction. They should be detectable out to a very high redshift
(Lamb \& Reichart 2000; Ciardi \& Loeb 2000; Gou et al. 2004). Hence,
GRBs are potentially a more promising ruler than SNe Ia at higher
redshifts.

This issue has attracted much attention in GRB community. Frail et
al. (2001) found that the geometrically-corrected gamma-ray energy
$E_{\rm{jet}}$ for long GRBs is narrowly clustered around $5\times
10^{50}$ ergs, suggesting that GRBs can be potentially a standard
candle. A refined analysis by Bloom, Frail, \& Kulkarni (2003a)
suggests that $E_{\rm{jet}}$ is clustered at $1.3\times 10^{51}$ ergs,
but the dispersion of $E_{\rm{jet}}$ is too large for the purpose of
constraining cosmological parameters.  Schaefer (2003) considered two
other luminosity indicators proposed earlier, i.e. the variability
(Fenimore \& Ramirez-Ruiz 2000; Reichart et al. 2001) and the spectral
lag (Norris, Marani, \& Bonnell 2000), for nine GRBs with known
redshifts, and pose an upper limit of
$\Omega_{\rm{M}}<0.35$ ($1\sigma$) for a flat universe. Using 12 {\em
BeppoSAX} bursts, Amati et al. (2002) found a relationship between the
isotropic-equivalent energy radiated during the prompt phase
($E_{\gamma,{{\rm{iso}}}}$) and rest frame peak energy in the GRB
spectrum ($E'_{\rm {p}}$), i.e., $E'_{\rm {p}}\propto
E_{\gamma,{{\rm{iso}}}}^{1/2}$.  This relation was confirmed and
extended to X-ray flashes by {\rm HETE-2} observations (Sakamoto et
al. 2004a; Lamb et al. 2005a). In addition, it also exists in the BATSE
bursts (Lloyd et al. 2000), and even in different pulses within a
single GRB (Liang, Dai \& Wu 2004).  Possible theoretical explanations
of this correlation have been proposed (Zhang \& M\'esz\'aros 2002a;
Dai \& Lu 2002; Yamazaki, Ioka, \& Nakamura 2004; Eichler \& Levinson
2004; Rees \& M\'esz\'aros 2005). Because of a large dispersion, this
relationship is not tight enough to serve a standard candle for
precision cosmology, either.

Ghirlanda et al. (2004a) found a tighter correlation between GRB jet
energy and $E_p^{'}$, which reads $E_{\rm{jet}}\propto (E'_p)^{3/2}$,
where $E_{\rm jet}= E_{\gamma,{\rm iso}} (1-\cos\theta_{\rm jet})$, and
$\theta_{\rm jet}$ is the jet opening angle inferred from the ``jet''
break time imprinted in the light curves (usually in the optical band,
and in some cases in the X-ray and the radio bands) by assuming a
uniform top-hat jet configuration. It is puzzling from the theoretical
point of view how a global geometric quantity (jet angle) would
conspire with $E_{\gamma,{\rm iso}}$ to affect $E'_p$. Nonetheless,
the correlation has a very small scatter which is arguably fine enough
to study cosmology.  By assuming that the correlation is intrinsic,
Dai, Liang, \& Xu (2004) constrained the mass content of the universe
to be $\Omega_{\rm{M}}=0.35\pm 0.15$ in the case of a flat Universe
with a sample of 14 GRBs.  They also constrained the dark matter
equation-of-state parameter in the range of $w=-0.84^{+0.57}_{-0.83}$
at $1\sigma$ level. Ghirlanda et al. (2004b) evaluated the goodness of
this relationship in different cosmologies by exploring the full
cosmological parameter space and came up with similar
conclusions. Friedmann \& Bloom (2005) suggested that this
relationship is only marginal but not adequate enough for a precision
cosmology study. The main criticisms are related to several assumptions
involved in the current Ghirlanda-relation, such as constant medium
density (which could vary in different bursts, e.g. Panaitescu \&
Kumar 2002), constant radiative efficiency (which also varies from
burst to burst, e.g. Lloyd-Ronning \& Zhang 2004; Bloom et al. 2003a
and references therein), and the assumption of the top-hat jet
configuration (in principle jets are possibly structured, Rossi et
al. 2002; Zhang \& M\'esz\'aros 2002b). Nonetheless, the
Ghirlanda-relation has motivated much work on measuring cosmology with
GRBs (e.g. Firmani et al. 2005; Qin et al. 2005; Xu et al. 2005; Xu
2005; Mortsell \& Sollerman 2005).

In this work, we further address the GRB standard candle problem by a
new statistical approach. Instead of sticking to the jet model and
searching for the correlation between $E'_{\rm p}$ and $E_{\rm jet}$
(which requires a model- and parameter-dependent jet angle), we start
with pure observable quantities to search for possible multi-variable
correlations by using a regression method. Similar technique was
employed by Schaefer (2003). The motivations of our analysis are
two-fold. First, within the jet model, there is no confident
interpretation to the Ghirlanda relation. It is relatively easy to
imagine possible correlations between $E'_{\rm p}$ and $E_{\gamma,{\rm
iso}}$ (e.g.  Zhang \& M\'esz\'aros 2002a; Rees \& M\'esz\'aros 2005),
since the latter is also a manifestation of the energy per solid angle
along the line of sight, which could be possibly related to the
emission spectrum. However, it is hard to imagine how the global
geometry of the emitter would influence the local emission
property\footnote{The simple Ghirlanda-relation could be derived from
the standard afterglow model and the Amati-relation, but one has to
assume that $t'_{\rm b}$ is constant for all GRBs, which is not true
(see also Wu, Dai, \& Liang 2004).}. Since there is no straightforward
explanation for the Ghirlanda relation, one does not have to stick to
this theoretical framework, but should rather try to look for some
empirical correlations instead. This would allow more freedom for possible
interpretations. Second, within various theoretical models (e.g.
Table 1 of Zhang \& M\'esz\'aros 2002a), the value of $E'_{\rm p}$
depends on multiple parameters. The problem is intrinsically
multi-dimensional. It is pertinent to search for multi-variable
correlations rather than searching for correlations between two
parameters only. The Ghirlanda-relation is a relation that bridges the
prompt emission and the afterglow phases. It is also worth checking
whether or not there are similar relationships for other
parameters. Below we perform a blind search for the possible
multi-variable correlations among several essential observable
quantities, including the isotropic gamma-ray energy
$E_{\gamma,{\rm{iso}}}$, the isotropic X-ray afterglow energy $E_{{\rm
{XA}},{\rm{iso}}}$, the isotropic optical afterglow energy
$E_{{\rm {OA}},{\rm{iso}}}$, the cosmological rest frame peak energy
$E_{\rm {p}}^{'}$, and the cosmological rest frame temporal break in
the {\em optical} afterglow light
curve ($t_{\rm{b}}^{'}$). We describe our sample selection criteria
and the data reduction method in \S 2.  Results of multiple regression
analysis are presented in \S3. A strong dependence of $E_{\gamma, {\rm
iso}}$ on $E^{'}_{\rm p}$ and $t_b^{'}$ is derived from our
multi-variable regression analysis. Regarding the $E_{\gamma,
{\rm{iso}}}(E_{\rm {p}}^{'}, t_{\rm{b}}^{'}$) relationship as a
luminosity indicator, in \S4 we explore the possible constraints on
cosmological parameters using the GRB sample.  In addition (\S5), we
perform Monte-Carlo simulations to investigate the characteristics of
the GRB sample satisfying the relationship observationally, and
examine how both the sample size and the observational precision
affect the constraints on
cosmological parameters. Conclusions and discussion are presented in
\S6. Throughout the work the Hubble constant is adopted as $H_0=71.3$
km s$^{-1}$ Mpc$^{-1}$.

\section{Sample selection and data reduction}
Our sample includes 15 bursts with measurements of the redshift $z$, the spectral peak
energy $E_p$, and the optical break time $t_b$. It has been suggested that the observed
Amati-relationship and the Ghirlanda-relationship are likely due to some selection
effects (Nakar \& Piran 2005; Band \& Preece 2005). The sample from which the relations
are drawn may therefore be ill-defined if the parent sample is the whole GRB population.
However, we believe that due to the great diversity of GRBs and their afterglow
observations, one does not have to require all GRBs to form a global sample to serve as a
standard candle. If one can identify a subclass of GRBs to act as a standard candle (such
as SNe Ia in the supernova zoo), such a sample could give meaningful implications to
cosmology.  Our selected GRBs belong to such a category, which assemble a unique and
homogeneous subclass. Since not all GRBs necessarily have an $E_p$ or a $t_b$, the parent
sample of our small sample is also only a sub-class of the whole GRB population. Notice
that in order to preserve homogeneity, we do not include those bursts whose afterglow
break times were observed in the radio band (GRB 970508, GRB 000418, GRB 020124) or in
the X-ray band (GRB970828), but were not seen in the optical band. Since we are not
sticking to the jet model, we do not automatically accept that there should be a temporal
break as well in the optical band.  We also exclude those bursts whose $E_p$ or $t_b$ are
not directly measured (but with upper or lower limits inferred from theoretical
modeling). This gives a sample of 15 bursts up to Feb, 2005. They are tabulated in Table
1 with the following headings: (1) the GRB name; (2) the redshift; the spectral fitting
parameters including (3) the spectral peak energy $E_{\rm {p}}$ (with error
$\sigma_{E_{\rm {p}}}$), (4) the low-energy spectral index $\alpha$, and (5) the
high-energy spectral index $\beta$; (6) the $\gamma$-ray fluence ($S_{\gamma}$)
normalized to a standard band pass ($1-10^4$ keV in the cosmological rest frame)
according to spectral fitting parameters (with error $(\sigma_{S_{\gamma}})$); (7) the
corresponding observation energy band; and (8) the references for these observational
data. Our GRB sample essentially resemble those used in Ghirlanda et al. (2004a), Dai et
al. (2004), and Xu et al. (2005). These bursts are included in the Table 1 of Friedmann
\& Bloom (2005), but that table also includes those bursts with only limits for $E_p$,
$t_j$, and $z$, as well as those bursts whose $t_b$ were observed in the non-optical
bands (or inferred from theoretical model fittings). We believe that our sample is more
homogeneous than the sample listed in Friedmann \& Bloom (2005).

The X-ray and optical afterglow data of
these GRBs are listed in Table 2 with the following headings: (1) the
GRB name; (2) the X-ray afterglow temporal decay index, $\alpha_x$;
(3) the epoch of x-ray afterglow observation (in units of hours); (4)
the 2-10 keV X-ray flux ($F_x$, in units of $10^{-13}$ erg cm $^{-2}$ s
$^{-1}$) at the corresponding epoch; (5) the 2-10 keV
X-ray afterglow flux normalized to 10 hours after the burst trigger
(including the error); (6) the temporal break (including the error) of
the optical afterglow light curves ($t_{\rm{b}}$); (7) the optical
temporal decay index before the break ($\alpha_1$); (8) the optical
temporal decay index after the break ($\alpha_2$); (9) the references;
(10) the R-band optical afterglow magnitudes at 11
hours. We find that the mean values of $\alpha_x$, $\alpha_{1}$, and
$\alpha_{2}$ in our sample are $1.41$, $1.0$, and $2.0$,
respectively. For those bursts whose $\alpha_x$, $\alpha_{1}$, and
$\alpha_{2}$ values are not available, we take
these means in our calculation.

With the data collected in Tables 1 and 2, we calculate the total
isotropic emission energies in the gamma-ray prompt phase
($E_{\gamma,{\rm iso}}$), in the X-ray afterglow
($E_{{\rm{XA}},{\rm{iso}}}$), and in the optical afterglow (R band)
($E_{{\rm {OA}},{\rm{iso}}}$), i.e.
\begin{equation}\label{Eg}
E_{\gamma,{\rm{iso}}}=\frac{4\pi D^2_L(z)S_{\gamma}k}{1+z},
\end{equation}
\begin{equation}\label{Exa}
E_{{\rm XA},{\rm{iso}}}=\frac{4\pi
D^2_L(z)\int^{t_2}_{t_1}F_x(t_1)t^{\alpha_{x}} dt }{1+z},
\end{equation}
and
\begin{equation}\label{Eoa}
E_{{\rm OA},{\rm{iso}}}=\frac{4\pi
D^2_L(z)(\int^{t_{\rm{b}}}_{t_1}F_R({t_1})t^{\alpha_{1}} dt
+\int^{t_2}_{t_{\rm{b}}}F_R(t_{\rm{b}})t^{\alpha_{2}} dt)}{1+z}.
\end{equation}
Here $D_L(z)$ is the luminosity distance at the redshift $z$, $k$ is a
k-correction factor to correct the observed gamma-ray fluence at an
observed bandpass to a given bandpass in the cosmological rest frame
($1-10^4$ keV in this analysis), $t_1$ and $t_2$ are, respectively, the
starting and the ending times of the afterglow phase, $F_x$ is the
flux of the X-ray afterglow in the 2-10 keV band, and $F_R$ is the
flux of the optical afterglow in the R band ($F_R=\nu_R f_\nu$). Since
the very early afterglows might be significantly different from the
later afterglows, which were not directly detected for the GRBs in our
sample, we thus take $t_1=1$ hour. We also choose $t_2=30$ days. The
derived $E_{\gamma,\rm{iso}}$, $E_{\rm XA,{\rm{iso}}}$, and $E_{\rm
OA,{\rm{iso}}}$ are tabulated in Table 3.

\section{Multiple Variable Regression Analysis}
\begin{figure}
\plotone{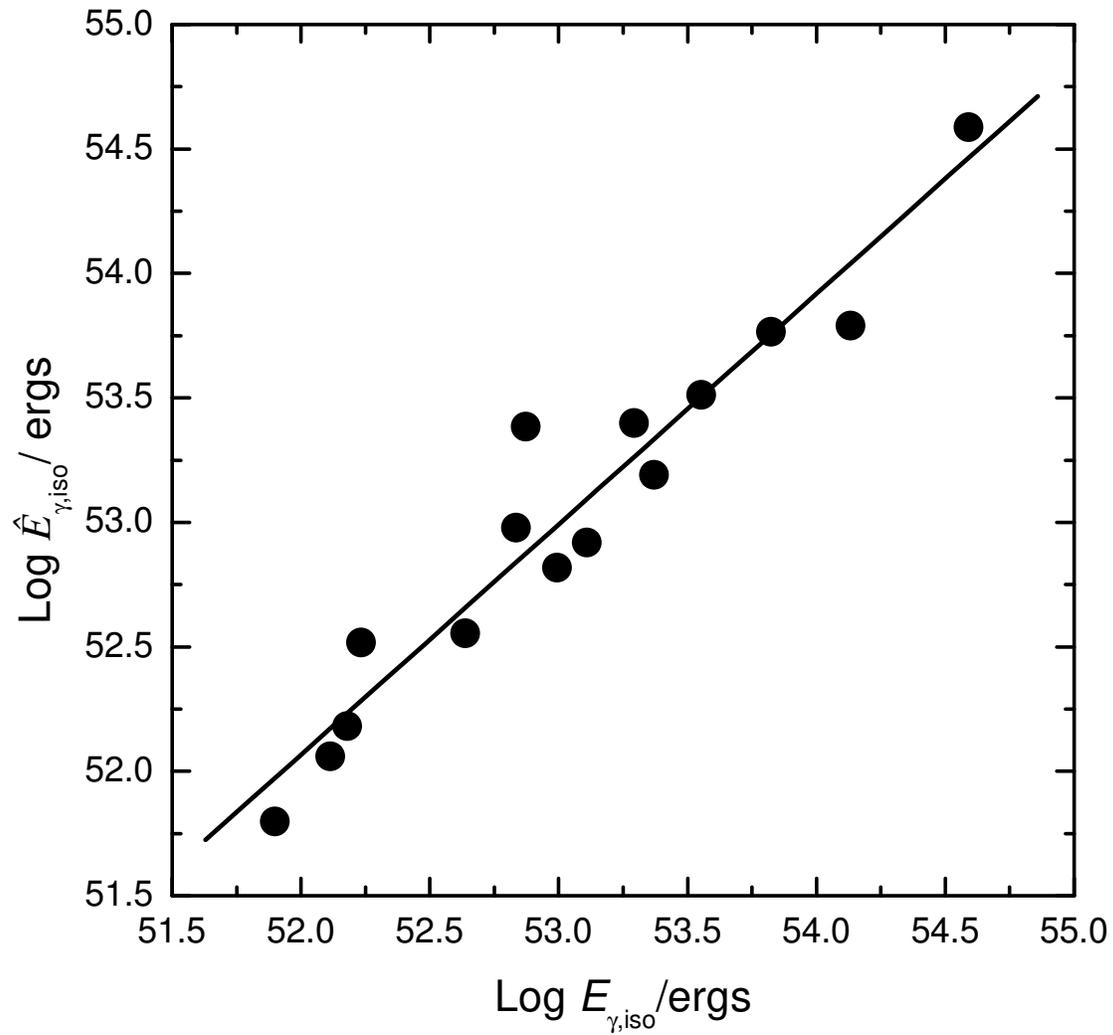} \caption{Log $\hat{E}_{\gamma, {\rm{iso}}}$ calculated by the empirical
relationship from our multiple variable regression analysis as compared with log
$E_{\gamma, {\rm{iso}}}$ derived from the observed fluence with the cosmological
parameters of $\Omega_M=0.28$ and $\Omega_{\Lambda}=0.72$. The solid line is the
regression line for the two quantities. \label{fig1}}
\end{figure}

\begin{figure}
\plotone{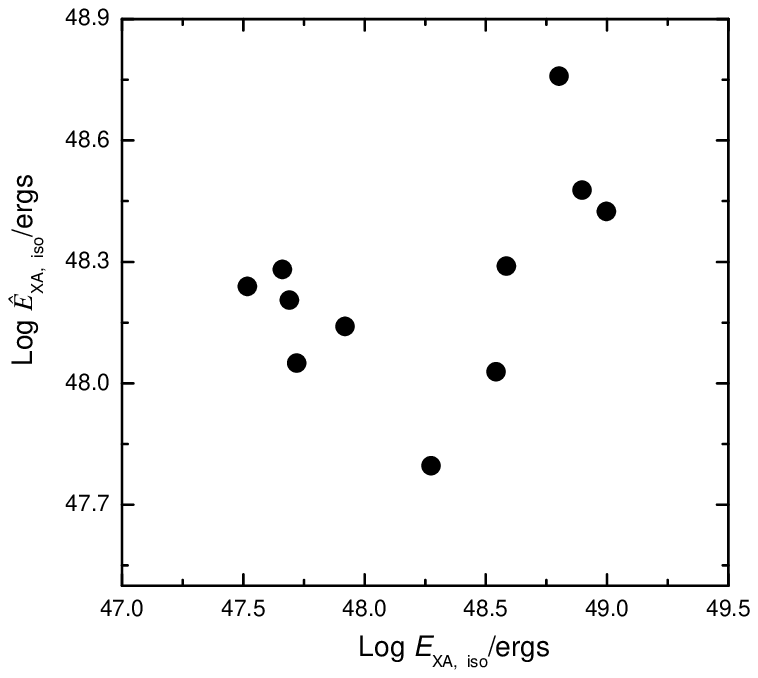} \caption{Log $\hat{E}_{{\rm XA}, {\rm{iso}}}$ calculated by the
empirical relationship from our multiple variable regression analysis as compared with
log $E_{{\rm XA}, {\rm{iso}}}$ derived with the cosmological parameters of
$\Omega_M=0.28$ and $\Omega_{\Lambda}=0.72$. \label{fig2}}
\end{figure}

\begin{figure}
\plotone{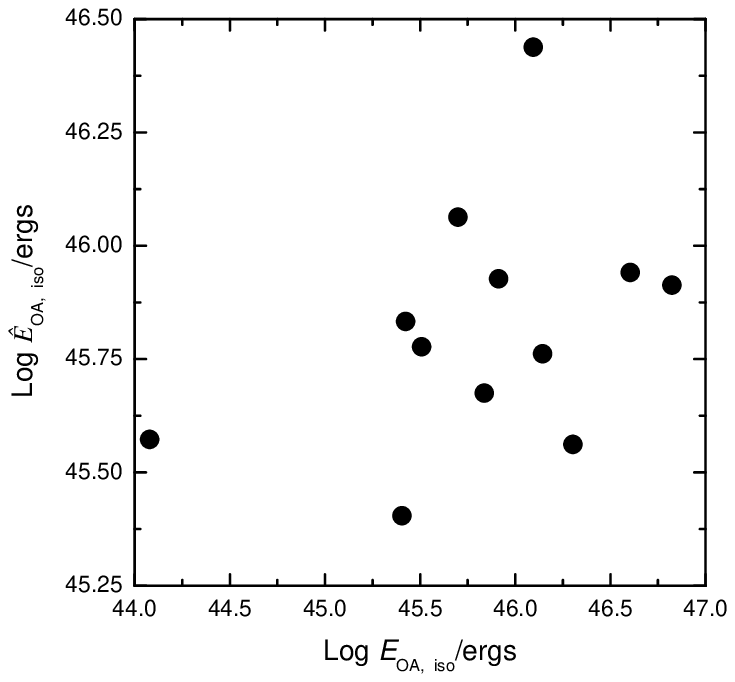} \caption{Log $\hat{E}_{{\rm OA}, {\rm{iso}}}$ calculated by the
empirical relationship from our multiple variable regression analysis as compared with
log $E_{{\rm OA}, {\rm{iso}}}$ derived with the cosmological parameters of
$\Omega_M=0.28$ and $\Omega_{\Lambda}=0.72$. \label{fig3}}
\end{figure}

As mentioned in \S1, previous authors interpret the
relationship among $E_{\gamma, {\rm{iso}}}$, $E_{\rm {p}}^{'}$, and
$t_{\rm{b}}^{'}$ based on the GRB jet model (e.g., Rhoads 1999; Sari
et al. 1999). In this scenario, the relationship among the three
quantities becomes the $E_{\rm{jet}} \propto (E'_{\rm {p}})^{3/2}$
relationship. When this relation is expanded, one gets $E_{\gamma,{\rm
iso}} t'_{\rm b}\propto (E'_{\rm p})^2$. The indices for
$E_{\gamma,{\rm iso}}$ and $t'_{\rm{b}}$ are not independent and are
bound by the jet model. However, since the current jet model is
difficult to accommodate the $E_{\rm{jet}}-E_{\rm {p}}$ relationship,
we no longer need to assume an underlying correlation between
$E_{\gamma,{\rm iso}}$ and $t'_{\rm{b}}$. We therefore leave all the
indices as free parameters and perform a multiple variable regression
analysis to search a possible empirical relationship among $E_{\gamma,
{\rm{iso}}}$, $E_{\rm {p}}^{'}$, and $t_{\rm{b}}^{'}$. We also extend
our analysis to search the dependence of $E_{X,{\rm{iso}}}$ and
$E_{R,{\rm{iso}}}$ on $E^{'}_{\rm {p}}$ and $t^{'}_{\rm{b}}$,
respectively. The regression model we use reads
\begin{equation}\label{MVR}
\hat{E}_{{\rm{iso}}}=10^{\kappa_{0}} E_{p}^{'\kappa_{1}} t_{b}^{' \kappa_{2}}
\end{equation}
where $E_{p}^{'}=E_{\rm {p}}(1+z)$ and
$t_{b}^{'}=t_{\rm{b}}/(1+z)$. We measure the significance level of the
dependence of each variable on the model by the probability of a
t-test ($p_t$). The significance of the global regression is measured
by a F-test (the $F$-test statistics and the corresponding
significance level $p_F$) and a Spearman linear correlation between $\log
\hat{E}_{{\rm{iso}}}$ and $\log {E}_{{\rm{iso}}}$ (the correlation
coefficient $r$ and the significance level $p_S$). We find that
$E_{\gamma, {\rm{iso}}}$ strongly depends on both $E_{\rm {p}}^{'}$
and $t_{\rm{b}}^{'}$ with a very small uncertainty (Table 4,
Fig.\ref{fig1}). The actual dependence format depends on the cosmology
adopted. For a flat Universe with $\Omega_M=0.28$, this relation reads
\begin{eqnarray}\label{MR}
\hat{E}_{\gamma,{\rm{iso,52}}}&=&(0.85\pm0.21)\times \left(\frac{E^{'}_{\rm p}}{100 ~{\rm
keV}}\right)^{1.94\pm0.17}\\
&&\times  \left(\frac{t'_{\rm b}}{1~{\rm day}}\right)^{-1.24\pm0.23},
\end{eqnarray}
where $\hat{E}_{\gamma,{\rm{iso,52}}}=\hat{E}_{\gamma,
{\rm{iso}}}/10^{52} {\rm ergs}$. However, when we test the possible
correlations among $E_{X, {\rm{iso}}}$ (or $E_{O, {\rm{iso}}}$),
$E_{\rm {p}}^{'}$ and $t_{\rm{b}}^{'}$, no significant correlation is
found (Table 4, Figs.\ref{fig2} and \ref{fig3}).

\section{Luminosity Indicator and Cosmological Implications}
The dispersion of the $\hat{E}_{\gamma,\rm{iso}}(E_{p}^{'},t_{b}^{'})$
relationship is so small that it could potentially serve as a
luminosity indicator for the cosmological study. This relationship is
{\em purely empirical}, exclusively using directly measured
quantities, and without
imposing any theoretical models and assumptions. It therefore suffers
less uncertainties/criticisms than does the Ghirlanda relation
(e.g. Friedmann \& Bloom 2005). Below we will discuss the cosmological
implications for this new empirical luminosity indicator.

The distance modulus of a GRB, which is defined as $\mu\equiv 5\log
(D_L/10 {\rm pc})$, could be measured by this luminosity indicator as
\begin{eqnarray}\label{mu}
\hat{\mu}&=&2.5[\kappa_0+\kappa_1\log E_{p}^{'}+\kappa_2\log t_{b}^{'}\\
&& -\log ({4\pi S_{\gamma}k})+\log ({1+z})]-97.45.
\end{eqnarray}
Since the luminosity indicator is cosmology-dependent, $\hat{\mu}$ is
also cosmology-dependent. We therefore cannot directly use this
relationship for our purpose. Ideally, it should be calibrated by local
GRBs (e.g., $z<0.1$), as is the case of Type Ia supernova
cosmology. However, the GRB low redshift sample is small. More
importantly, the local GRBs appear to have different characteristics
than the cosmological ones (e.g. long lag, less luminous etc), so that
they may not belong to the subclass of GRBs we are discussing. We are
left out without a {\em real} (cosmological-independent) luminosity
indicator at this time.

We adopt the following method to circumvent the difficulty. We first
recalibrate this relationship in each cosmological model, and then
calculate the goodness of the relationship in that cosmology by
$\chi^2$ statistics. We then construct a relation which is weighed by the
goodness of each cosmology-dependent relationship, and use this
cosmology-weighed relationship to measure the Universe. The procedure
to calculate the probability function of a cosmological parameter set
(denoted as $\Omega$, which includes both $\Omega_M$ and
$\Omega_\Lambda$) is the following.

(1) Calibrate and weigh the luminosity indicator in each
cosmology. Given a particular set of cosmological parameters
($\bar\Omega$), we perform a multiple variable regression analysis and
get a best-fit correlation $\hat{E}_{\gamma,\rm{iso}}(\bar{\Omega};
E_{p}^{'},t_{b}^{'})$. We evaluate the probability ($w(\bar{\Omega})$)
of using this relation as a cosmology-independent luminosity indicator
via $\chi^2$ statistics, i.e.
\begin{equation}\label{chir}
\chi^2_{w}(\bar{\Omega})=\sum_{i}^{N}\frac{[\log \hat{E}^{i}_{\gamma, {\rm iso}}(\bar{\Omega})-\log E^{i}_{\gamma, {\rm
iso}}(\bar{\Omega})]^2}{\sigma_{\log \hat{E}^{i}_{\gamma, {\rm
iso}}}^2(\bar{\Omega})}.
\end{equation}
The smaller the $\chi^2_{r}$, the better the fit, and hence, the
higher probability for this cosmology-dependent relationship to serve
as a cosmological independent luminosity indicator. We assume that the
distribution of the $\chi^2_{w}(\bar{\Omega})$ is normal, so that the
probability can be calculated as
\begin{equation}\label{weight}
w(\bar{\Omega})\propto e^{-\chi^2_{w}(\bar{\Omega})/2}.
\end{equation}

(2) Regard the $\hat{E}_{\gamma,\rm{iso}}(\bar{\Omega};
E_{p}^{'},t_{b}^{'})$ relationship derived in each cosmology as a
cosmology-independent luminosity indicator without considering its
systematic error, and calculate the corresponding distance modulus
$\hat{\mu}(\bar{\Omega})$ [eq. \ref{mu}] and its error
$\sigma_{\hat{\mu}}$, which is
\begin{eqnarray}\label{err2}
\sigma_{\hat{\mu}_{i}}&=&\frac{2.5}{\ln 10}[(\kappa_1 \frac{\sigma_{E_{\rm p,i}^{'}}}{
E_{\rm p,i}^{'}})^2+(\kappa_2\frac{\sigma_{t_{\rm b,i}^{'}}}{t_{\rm b,i}^{'}})^2
+(\frac{\sigma_{S_{\gamma,i}}}{S_{\gamma,i}})^2\\
&&+(\frac{\sigma_{k_i}}{k_i})^2+(\frac{\sigma_{z_i}}{1+z_{i}})^2]^{1/2}.
\end{eqnarray}

(3) Calculate the theoretical distance modulus $\mu (\Omega)$ in an
arbitrary set of cosmological parameters (denoted by $\Omega$), and
calculate the $\chi^2$ of $\mu (\Omega)$ against $\hat{\mu}(\Omega)$,
i.e.
\begin{equation}\label{chis}
\chi^2(\bar{\Omega}|\Omega)=\sum_{i}^{N} \frac{[\hat{\mu}_i(\bar{\Omega})-\mu_i(\Omega)]^2}{\sigma_{
\hat{\mu}_{i}}^2(\bar{\Omega})}
\end{equation}

(4) Assuming that the distribution of $\chi^2(\bar{\Omega}|\Omega)$ is
also normal, calculate the probability that the cosmology parameter
set $\Omega$ is the right one according to the luminosity indicator
derived from the cosmological parameter set $\bar{\Omega}$, i.e.
\begin{equation}\label{pp}
p(\bar{\Omega}|\Omega)\propto e^{ -\chi^2(\bar{\Omega}|\Omega)/2}.
\end{equation}
With eq.(\ref{weight}), we can define a cosmology-weighed likelihood by
$w(\bar{\Omega})p(\bar{\Omega}|\Omega)$.

(5) Integrate $\bar{\Omega}$ over the full cosmology parameter space
to get the final normalized probability that the cosmology $\Omega$ is
the right one, i.e.
\begin{equation}\label{PW}
p(\Omega)=\frac{\int_{\bar{\Omega}}w(\bar{\Omega})
p(\bar{\Omega}|\Omega)d\bar{\Omega}}{\int_{\bar{\Omega}}
w(\bar{\Omega})d\bar{\Omega}}.
\end{equation}

In our calculation, the integration in eq.(\ref{PW}) is computed
through summing over a wide range of the cosmology parameter space to
make the sum converge, i.e.,
\begin{equation}\label{PWsum}
p(\Omega)=\frac{\sum_{\bar{\Omega}_i}w(\bar{\Omega}_i)
p(\bar{\Omega}_i|\Omega)}{\sum_{\bar{\Omega}_i}w(\bar{\Omega}_i)}.
\end{equation}

\begin{figure}
\plotone{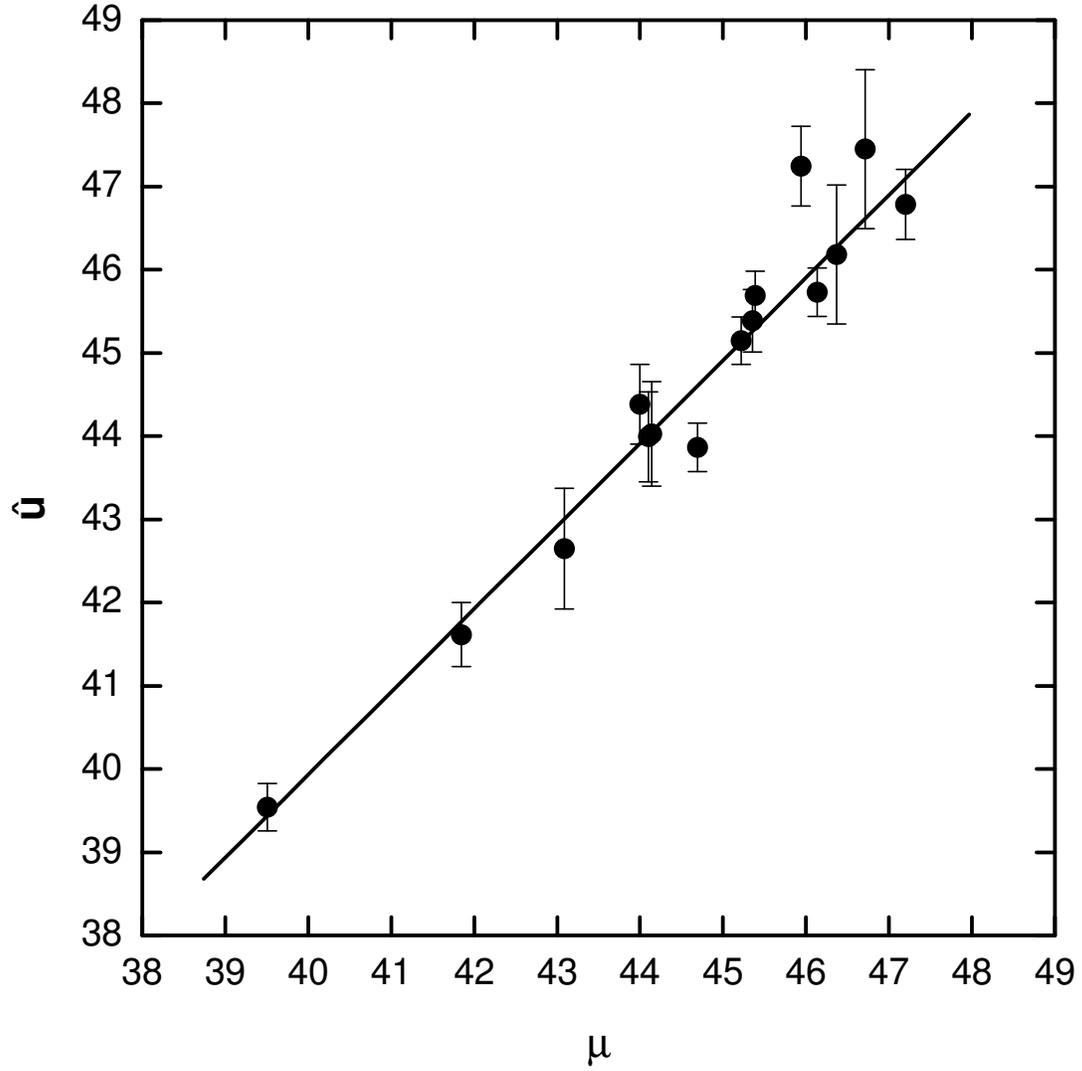} \caption{The distance modulus derived from the data, $\hat{\mu}$, and
its observational error, $\sigma_{\hat{\mu}}$, are plotted against the distance modulus
derived from theory, $\mu$. The cosmological parameters are adopted as $\Omega_M=0.28$
and $\Omega_{\Lambda}=0.72$. \label{fig4}}
\end{figure}

The essential ingredient of our method is that we do not include the
systematical error of the $\hat{E}_{\gamma,\rm{iso}}(\bar{\Omega};
E_{p}^{'},t_{b}^{'})$ relationship into
$\sigma_{\hat{\mu}_{s,i}}$. Instead, we evaluate the probability that
a particular relationship can be served as a cosmology-independent
luminosity indicator using its systematical error, and integrate over
the full cosmology parameter space to get the final probability of a
cosmology with the parameter set $\Omega$. In Figure \ref{fig4} we
plot $\hat{\mu}$ against $\mu$ with $\sigma_{\hat{\mu}_{s}}$ in the
case of $\Omega=0.28$ and $\Omega_{\Lambda}=0.72$ cosmology. Similar
investigation could be done for other cosmologies. Below, we will
apply the approach discussed above to investigate the possible
implications on cosmography and cosmological dynamics with our GRB
sample.

\subsection{Implications for $\Omega_{\rm{M}}$ and $\Omega_\Lambda$}

In a Friedmann-Robertson-Walker (FRW) cosmology with mass density
$\Omega_{\rm{M}}$ and vacuum energy density $\Omega_\Lambda$, the
luminosity distance is given by
\begin{eqnarray}\label{DL}
D_L & = & c(1+z)H_0^{-1}|\Omega_k|^{-1/2}{\rm
sinn}\{|\Omega_k|^{1/2}\nonumber \\ & & \times
\int_0^zdz[(1+z)^2(1+\Omega_{\rm{M}}z)-z(2+z)\Omega_\Lambda]^{-1/2}\},
\end{eqnarray}
where $c$ is the speed of light, $H_0$ is the present Hubble constant,
$\Omega_k=1-\Omega_{\rm{M}}-\Omega_{\Lambda}$ denotes the curvature of
the universe, and ``sinn" is sinh for $\Omega_k>0$ and sin for
$\Omega_k<0$. For a flat universe ($\Omega_k=0$), Eq.(\ref{DL}) turns
out to be $c(1+z)H_0^{-1}$ multiplies the integral. We calculate
$p(\Omega)$ with our GRB sample, where
$\Omega=(\Omega_{\rm{M}},\Omega_{\Lambda})$. Since both
$[{\sigma_z}/({1+z})]^2$ and $(\sigma_{k}/k)^2$ are significantly
smaller than the other terms in eq.(\ref{err2}), they are ignored in
our calculations.  Shown in Figure \ref{fig5} are the most possible
value of $(\Omega_{\rm{M}}, \Omega_{\Lambda})$ and the $1\sigma$ to
$3\sigma$ contours of the likelihood in the
($\Omega_{\rm{M}}$,$\Omega_\Lambda$) plane. The most possible value of
$(\Omega_{\rm{M}},\Omega_{\Lambda})$ is $(0.28, 0.64)$. The contours
show that $0.05<\Omega_{\rm{M}}<0.50$ at $1\sigma$, but
$\Omega_\Lambda$ is poorly constrained, i.e. $\Omega_{\Lambda}<1.2$ at
$1\sigma$. For a flat Universe, as denoted as the dashed line in
Figure \ref{fig5}, the constraints are tighter,
i.e. $0.13<\Omega_{\rm{M}}<0.49$ and $0.50<\Omega_{\Lambda}<0.85$ at
$1\sigma$.
\subsection{Implications for the cosmology dynamics}
Riess et al. (2004) found the evidence from SNe Ia data that the Universe
was switched from a past decelerating phase to the currently accelerating
phase at an epoch of $z_t = 0.46 \pm 0.13$, assuming that the
decelerating factor $q$ evolves with redshift as $q(z)=q_0+ z
dq/dz$. Following Riess et al. (2004), we also take $q(z)=q_0+ z dq/dz$ to
analyze the implications for $q_0$ and $dq/dz$ from the current GRB
sample. The luminosity distance in a $(q_0,dq/dz)$ model can be
written as
\begin{equation}\label{DL2}
D_L = \frac{c(1+z)}{H_0} \int_0^z e^{-\int_{0}^{u}
[1+q(u)]d\ln(1+u)}du.
\end{equation}
We then calculate the values of $P(\Omega)$ (where $\Omega=(q_0, dq/dz)$)
using the cosmology-weighed standard candle method discussed
above. Shown in Figure \ref{fig6} are the most possible values of
$(q_0,dq/dz)$ and their likelihood interval contours from $1\sigma$ to
$3\sigma$. The most possible values of $(q_0,dq/dz)$ are $(-1.0,
1.12)$, and at $1\sigma$ level their values are constrained in the
ranges of $-2.23<q_0<0.26$ and $-0.07<dq/dz<3.48$. Although the
current sample still does not place a tight constrain on both $q_0$
and $dq/dz$, it shows that $q_0$ tends to be less than 0 and $dq/dz$
tends to be greater than 0, suggesting that the Universe is
accelerating now. At a given epoch $z_t$ in the past, $q(z_t)=0$
should be satisfied, which denotes the transition between the past
decelerating phase and the currently accelerating phase. The likelihood
function of $z_t$ derived from the current GRB sample is shown in
Figure \ref{fig7}.  We calculate the best value of $z_t$ by
\begin{equation}\label{zt}
\hat{z}_t=\frac{\sum p(z_t)z_t}{\sum p(z_t)},
\end{equation}
and get $\hat{z}_t=0.78^{+0.32}_{-0.23}$ at $1\sigma$.

\section{Simulations}

We have shown that using the analysis method proposed in this paper, one can place some
constraints on the cosmology parameters with our GRB sample. These constraints are,
however, weaker than those obtained with the SNe Ia data, and they are of uncertainties
because of the small GRB sample effect. To increase the significance of the constraints,
one needs a larger sample and smaller error bars for the measurements. In order to access
the characteristics of the GRB sample satisfying our relationship observationally and how
the sample size and the measurement precision affect the standard analysis, we perform
some Monte Carlo simulations. We simulate a sample of $10^3$ GRBs. Each burst is
characterized by a set of parameters denoted as ($z$, $E_p$, $S_{\gamma}$, $t_b$). A
fluence threshold of $S_{th,\gamma}=10^{-7}$ erg s$^{-1}$ is adopted. Since the observed
$t_b$ is in the range of $0.4\sim 6$ days, we also require that $t_b$ is in the same
range to account for the selection effect to measure an optical lightcurve break. Our
simulation procedures are described as follows.

(1) Model the accumulative probability distributions of $E_p$, $E_{iso}$, and $z$ by the
observational data. We first obtain the differential distribution of these measurements.
The $E_p$ distribution is derived from the GRB spectral catalog presented by Preece et
al. (2000), which is well modeled by $dp/d\log E_{p}\propto \exp [-2(\log
E_{p,2}-0.38)^2/0.45^2]$, where $E_{p,2}=E_p/100$ (Liang, Dai, \& Wu 2004). The $E_{iso}$
distribution is obtained from the current sample of GRBs with known redshifts. Since the
$E_{iso}$ distribution suffers observational bias at the low $E_{iso}$ end, we consider
only those bursts with $E_{iso}>10^{51.5}$ ergs, and get $dp/d\log E_{iso}\propto -0.3
\log E_{iso}$\footnote{Our simulations do not sensitively depend on the $E_{iso}$
distribution. We have used a random distribution between $10^{51.5}\sim 10^{54.5}$ ergs,
and found that the characteristics of our simulated GRBs sample are not significantly
changed.}. The redshift distribution is derived by assuming that the GRB rate as a
function of redshift is proportional to the star formation rate. The model SF2 from
Porciani \& Madau 2001 is used in this analysis. We truncate the redshift distribution at
10. Based on these differential distributions, we obtain the accumulative distributions,
$p_x$, where $x$ is one of these parameters. We use the discrete forms of these
distributions to save the calculation time. The bin sizes of $\log E_p$, $\log E_{iso}$,
and $z$ are taken as 0.025, 0.1, and 0.01, respectively.

(2) Simulate a GRB. We first generate a random number $m$ ($0<m\leq
1$), and obtain the value of $x_m$ from the inverse function of
$p_x(x_m)=m$, i.e., $x_m=p_x^{-1}(m)$. Since $p_x$ is in a discrete
form, we search for a bin $i$, which satisfies $p_x(x_{i})<m$ and
$p_x(x_{i+1})>m$, and calculate the $x_m$ value by
$x_m=(x_{i+1}+x_{i})/2$. Repeating this step for each parameter, we
get a simulated GRB characterized by a set of parameters denoted as
($z$, $E_{iso}$, $E_p$).

(3) Calculate $S_{\gamma}$ and examine whether or not the $S_{\gamma}$
satisfies our threshold setting. The gamma-ray fluence is calculated
by $S=E_{\rm{iso}}(1+z)/4\pi D_{L}^{2}(z)$, where $D_{L}(z)$ is the
luminosity distance at $z$ (for a flat universe with $\Omega_M=0.3$).
If $S<S_{th, \gamma}$, the burst is excluded.

(4) Derive $t_b$. We first infer a $t_b$ value from our empirical
relation in a flat Universe of $\Omega_M=0.3$, then assign a deviation
($\Delta t_b$) to the $t_b$ value.  The distribution of $\Delta t_b$
is taken as $d N/d \Delta \log t_b=\exp(-\Delta t_b^2/2\sigma)$, where
$\sigma=0.1$. This typical value is taken according to the current
sample, which gives the mean and medium deviations as $\sigma=0.15$
and $\sigma = 0.11$, respectively.
If the $t_b$ value is in the range of $0.4<t_b<6 $ days,
this burst is included in our sample. Otherwise, it is excluded.

(5) Assign {\em observational errors} to $E_p$, $S_{\gamma}$, and
$t_b$. Since the observed $\sigma_x/x$ is about $10\%-20\%$, we take
the errors as $\sigma_x/x=0.25k$ with a lower limit of
$\sigma_x/x>5\%$, where $k$ is a random number between $0\sim 1$.

(6) Repeat steps (2) and (5) to obtain a sample of $10^{3}$ GRBs.

The distributions of $z$, $E_{p}$, and $S_{\gamma}$ for the simulated
GRB sample are shown in Figure \ref{fig8} (solid lines). The observed
distributions of these quantities are also shown for comparison
(dotted lines). The observed redshift distribution is
derived from the current GRB sample with known redshifts (45 GRBs).
The observed $E_p$ distribution is taken from Preece et al. (2000).
The observed $S_{\gamma}$ is derived from the BATSE Current GRB
sample\footnote{http://cossc.gsfc.nasa.gov/batse/} (Cui, Liang, \& Lu
2005). The comparisons indicate that the mock GRB sample tends to be a
softer (low $E_p$) and brighter (high $S_{\gamma}$) one. The
redshifts of the mock GRB sample tend to be higher than the
current GRB sample, but this might be due to observational biases
against high redshift GRBs (Bloom 2003).

We investigate the effect of the sample size on the cosmological constraints with our
mock GRB sample. We randomly select sub-samples of 25, 50, 75, and 100 GRBs from the mock
GRB sample. We compare the $1\sigma$ contours of likelihood distributions in the
($\Omega_{\rm{M}}$, $\Omega_\Lambda$)-plane derived from these sub-samples in the left
panel of Figure \ref{fig9}. It is evident that, as the sample size increases, the
constraint on $\Omega_{\rm{M}}$ and $\Omega_\Lambda$ becomes more tighter.  Comparing the
left panel of Figure \ref{fig9} with the Figure 8 in Riees et al. (2004), we find that
the likelihood contour derived from the sub-sample of 50 GRBs is comparable to that
derived from the gold sample of 157 SNe Ia.

Precision cosmology requires accurate observations. Modern
sophisticated observation techniques in distant SNe Ia (e.g. Riess et
al. 1998; Schmidt et al. 1998; Perlmutter et al. 1999) and cosmic
microwave background (CMB) fluctuations (e.g. Bennett et al. 2003;
Spergel et al. 2003) have made great progress on modern precision
cosmology. We inspect the uncertainties of the distance modulus
derived from the SNe Ia data, and find that the average uncertainty
is $\bar{\sigma}_{DM}\sim 0.25$, while for our GRB sample it is
0.45. Increasing observational precision (i.e. reducing the errors)
should significantly improve the constraints on the cosmological
parameters. We simulate another GRB sample with systematically smaller
{\em observational errors}, i.e. $\sigma_x/x=0.15k$ in the step 5 of
our simulation procedure.  We get a sample
with $\bar{\sigma}_{DM}\sim 0.28$, comparable to the SNe Ia gold
sample. The comparison of the likelihood contours ($1\sigma$) in the
$\Omega_{\rm{M}}-\Omega_\Lambda$ plane derived from a sample of 50
mock GRBs with $\bar{\sigma}_{DM}\sim 0.45$ (the line contour) and
with $\bar{\sigma}_{DM}\sim 0.28$ (the grey contour) are shown in the
right panel of Figure \ref{fig9}. It is found that the latter is
significantly tighter, which is comparable to that derived from a
sample of 100 mock GRBs with an average error in modulus of 0.45.

The results in Figures \ref{fig9} indicate that tighter constraints on
cosmological parameters could be achieved by either enlarging the
sample size or increasing the observational precision. If a
sample of 50 GRBs with a comparable observational precision as SNe Ia
gold sample could be established, the constraints are even tighter
than those derived from the SNe Ia gold sample.

\section{Conclusions and Discussion}
Without imposing any theoretical models and assumptions, we
investigate the relationship among $E_{\gamma,{\rm{iso}}}$, $E_{\rm
{p}}^{'}$ and $t_{\rm{b}}^{'}$ using a multiple variable regression
method. Our GRB sample includes 15 bursts, whose $E_{\rm {p}}^{'}$ and
$t_{\rm{b}}^{'}$ are well measured. The results indicate that $E_{\gamma,
{\rm{iso}}}$ strongly depends on both ${E}_{\rm {p}}^{'}$ and
$t_{\rm{b}}^{'}$ with a very small dispersion, e.g. eq.(\ref{MR}) for
a flat Universe with $\Omega_M=0.28$. We also perform a similar
analysis by replacing $E_{\gamma,{\rm iso}}$ by the isotropic
afterglow energies in the X-ray and optical bands, and find that these
energies are essentially not related to $E_{\rm {p}}^{'}$ and
$t_{\rm{b}}^{'}$ at all. We then use the $E_{\gamma,
{\rm{iso}}}(E_{\rm {p}}^{'}, t_{\rm{b}}^{'}$) relationship as a
luminosity indicator to infer the possible cosmology implications from
the GRB sample. Since this relationship is cosmology-dependent, we
suggest a new method to weigh various cosmology-dependent
relationships with its probability of being the right one, and use the
cosmology-weighed standard candle to explore the most plausible
cosmological parameters. Our results show that the most possible
values are $(\Omega_{\rm{M}}, \Omega_\Lambda)=(0.28,0.64)$. At
$1\sigma$ level, we have $0.05<\Omega_{\rm{M}}<0.50$ and
$\Omega_{\Lambda}<1.2$. In the case of a flat Universe, the $1\sigma$
constraints are $0.13<\Omega_{\rm{M}}<0.49$ and
$0.50<\Omega_{\Lambda}<0.85$. The decelerating factor of the Universe
($q$) and its cosmological evolution $(dq/dz)$ are also investigated
with an evolutionary form of $q=q_0+zdq/dz$. The GRB sample implies
that the most possible values of $(q_0,dq/dz)$ are $(-1.00, 1.12)$,
and they are constrained in the ranges of $-2.23<q_0<0.26$ and
$-0.07<dq/dz<3.48$ at $1\sigma$ level. A transition redshift between
the deceleration and the acceleration phases of the Universe is
inferred as $\hat{z}_t=0.78^{+0.32}_{-0.23}$ at $1\sigma$ level from
the GRB sample.

\begin{figure}
\plotone{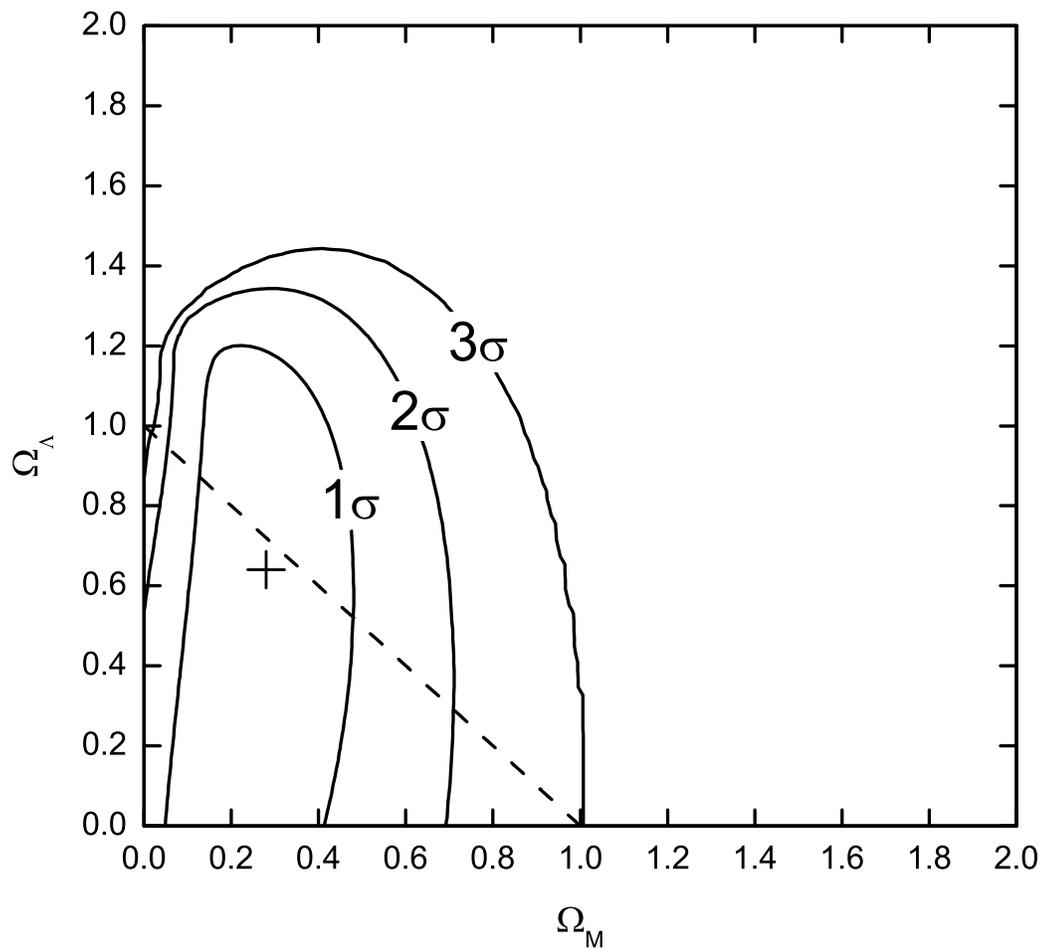} \caption{The contours of likelihood interval distributions in the
($\Omega_{\rm{M}}$, $\Omega_\Lambda$)-plane inferred from the current GRB sample using
the method developed in \S4. The cross marks the most possible value of
$(\Omega_{\rm{M}},\Omega_{\Lambda})$, which is $(0.28, 0.64)$. The contours give
$0.05<\Omega_{\rm{M}}<0.50$($1\sigma$). Considering a flat universe (the dashed line),
the contours yield $0.13<\Omega_{\rm{M}}<0.49$ and $0.50<\Omega_{\Lambda}<0.85$
($1\sigma$). \label{fig5}}
\end{figure}

\begin{figure}
\plotone{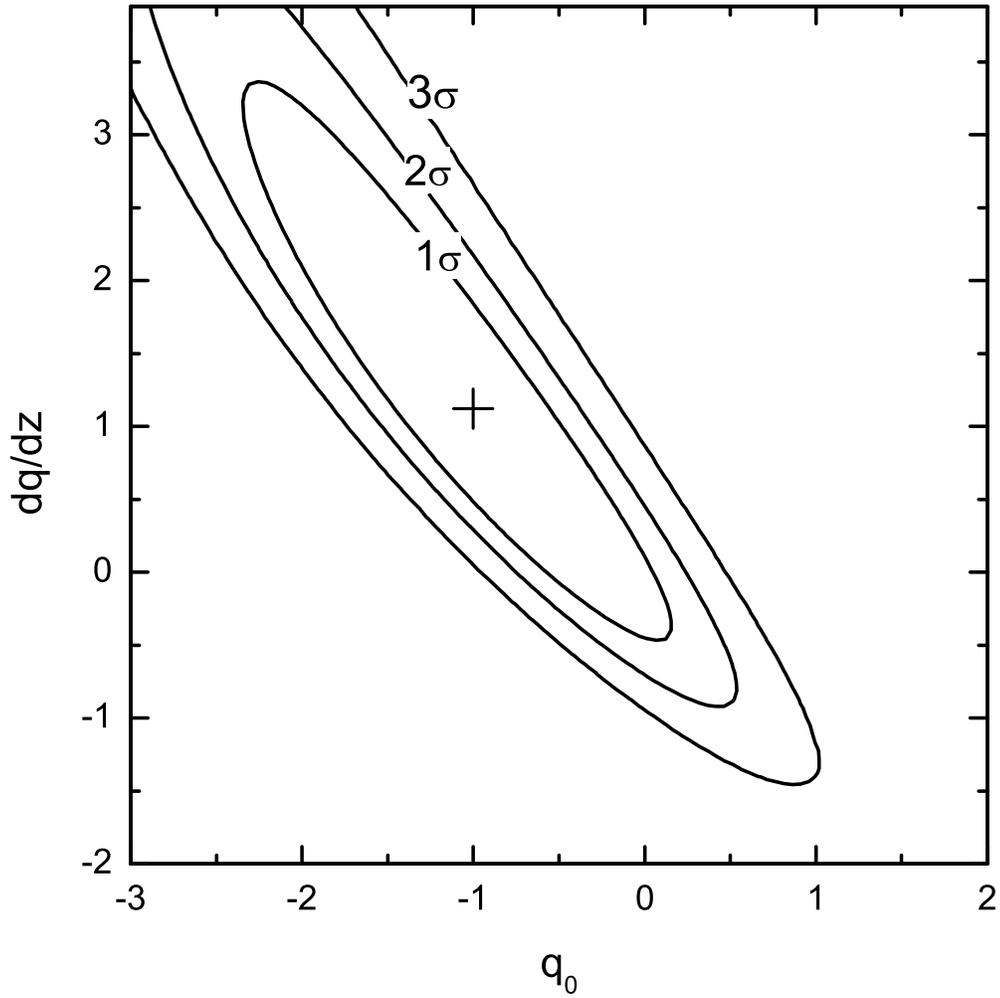} \caption{The contours of likelihood interval
distributions in the $(q_0,dq/dz)$-plane
inferred from the current GRB sample using the method developed in
\S4. The most possible values of $(q_0, dq/dz)$ are $(-1.00, 1.12)$
(the cross). At $1\sigma$ level their values are constrained in the
ranges of $-2.23<q_0<0.26$
and $-0.07<dq/dz<3.48$, respectively. \label{fig6}}
\end{figure}

\begin{figure}
\plotone{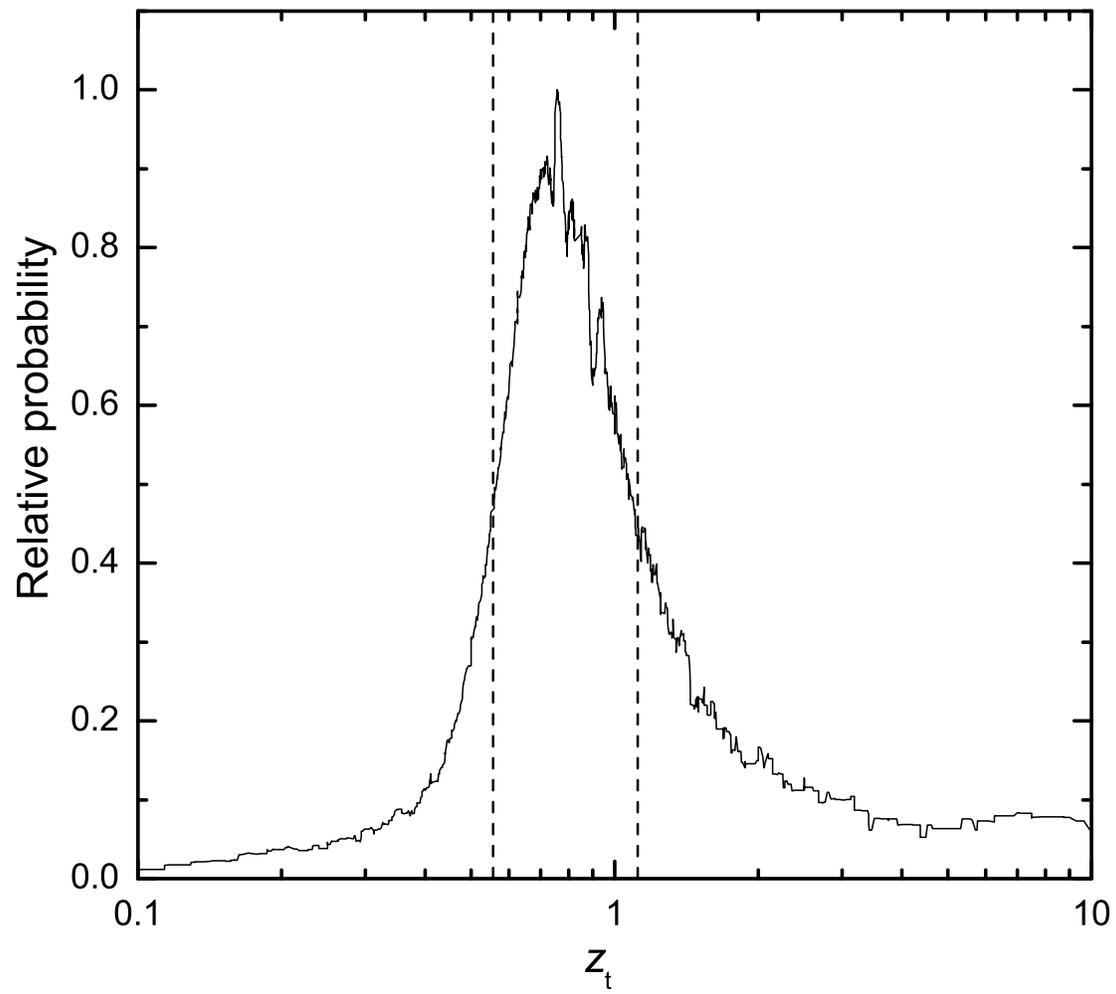} \caption{The smoothed likelihood function of the
transition redshift from a decelerating Universe to
an accelerating Universe inferred from the current GRB sample. The
dashed lines mark the $1\sigma$ region, and the best value of
$\hat{z}_t$ is $0.78^{+0.32}_{-0.23}$. \label{fig7}}
\end{figure}

As a luminosity indicator, our model-independent $E_{\rm \gamma,iso}
(E^{'}_p, t^{'}_b)$ relationship takes the advantage upon the previous
Ghirlanda-relation in that only pure observational data are
involved. Since this luminosity
indicator is cosmology-dependent, we use a strategy through weighing
this relationship in all possible cosmologies to statistically study
the cosmography and cosmological dynamics. A similar method has been
used in the SN cosmology when dealing with the uncertainty of the
present Hubble constant $H_0$. In their method (e.g. Riess et
al. 1998), the systematic error of $H_0$ is not included to calculate
the error of the distance modulus. Rather, they integrated the
probability of $H_0$ over a large range values (without weighing for
each value of $H_0$). This is the so-called marginalization method. We
also perform this marginalization method to deal with our coefficients
($\kappa_0$, $\kappa_1$ and $\kappa_2$), and re-do the
cosmology-analysis. This is equivalent to integrating over the whole
cosmology parameter space without weighing, i.e.,
\begin{equation}\label{PM}
p(\Omega)=\int_{\bar{\Omega}}p(\bar{\Omega}|\Omega)d\bar{\Omega}.
\end{equation}
The result using this method to constrain $\Omega_{\rm{M}}$ and
$\Omega_{\Lambda}$ is presented in Figure \ref{fig10}.  Comparing it
with Fig.\ref{fig5}, we find that both methods give consistent
results, but Fig.\ref{fig5} gives a tighter constraint on cosmological
parameters. This is understandable, since the weighing method reduces
the contributions of side lobe around the ``true'' cosmologies. In any
case, an essential ingredient of both methods is that the uncertainty
of the standard candle itself is not included in calculating the
uncertainty of the distance modulus derived from the data. If the
uncertainty of the standard candle is indeed included in the
uncertainty of the distance modulus, with eqs.(\ref{PW}) and
(\ref{PM}) to calculate $p(\Omega)$, one gets a very loose constraint
(Fig.\ref{fig11}). Even at $1\sigma$ level the current GRB sample
cannot place any meaningful constraints on both $\Omega_{\rm{M}}$ and
$\Omega_{\Lambda})$. We believe, however, that in such a treatment,
the uncertainty of the distance modulus is over-estimated, since the
error introduced from measurements should not be mixed with the
systematic uncertainty of the standard candle.

The GRB sample from which our relationship is drawn is currently
small. The constraints on the cosmological parameters derived from
this sample are weaker than those from the SNe Ia gold sample.
Our simulations indicate that either the enlargement of the sample
size or the increase of the observational precision could greatly
improve the constraints on the cosmological parameters. A sample of 50
bursts with the current observational precision would be comparable to
the 157 SNe Ia gold sample in constraining cosmology, and a better
constraint is achievable with better observational precisions or an
even larger sample size.

Our simulations also indicate that the GRB sample satisfying our relationship
observationally tends to be a soft and bright one, for which $t^{'}_b$ is in the
reasonable range for detection. Detailed optical afterglow light curves covering from a
few hours to about ten days after the burst trigger\footnote{Starting from about ten
days, the contributions from the underlying SN and host galaxy components may become
prominent, and the afterglow level may be too faint to be detected.} are required to
measure the $t_b$ value. The observed $t_b$ ranges from 0.4 days to 5 days in the current
GRB sample. In the CGRO/BATSE duration
table\footnote{http://gammaray.msfc.nasa.gov/batse/grb/catalog/current}, there are $\sim
1500$ long GRBs. To test the probability of a BATSE burst having a $t_b$ in the range of
0.4 days to 5 days, we perform a simulation similar to that described in \S 5, but take
the $E_p$ and $S_{\gamma}$ distributions directly from the BATSE observations. We find
that the probability is $\sim 0.15$ in the cosmology of $\Omega_M=0.3$ and
$\Omega_\Lambda=0.7$. Among well localized GRBs about $50\%$ of bursts are optically
bright. We thus estimate that there have been $\sim 110$ BATSE GRBs that might have been
detected to satisfy our $E_{\rm \gamma,iso} (E^{'}_p, t^{'}_b)$ relationship. As shown in
Figure \ref{fig9}, such a sample is comparable to the SNe Ia gold sample for constraining
cosmological parameters. A dedicated GRB mission carrying a BATSE-liked GRB detector and
having the capability of precisely localizing and following up GRBs (like {\em Swift})
would be ideal to establish a homogenous GRB sample to perform precision GRB cosmology
(Lamb et al. 2005b).

\begin{figure}
\plotone{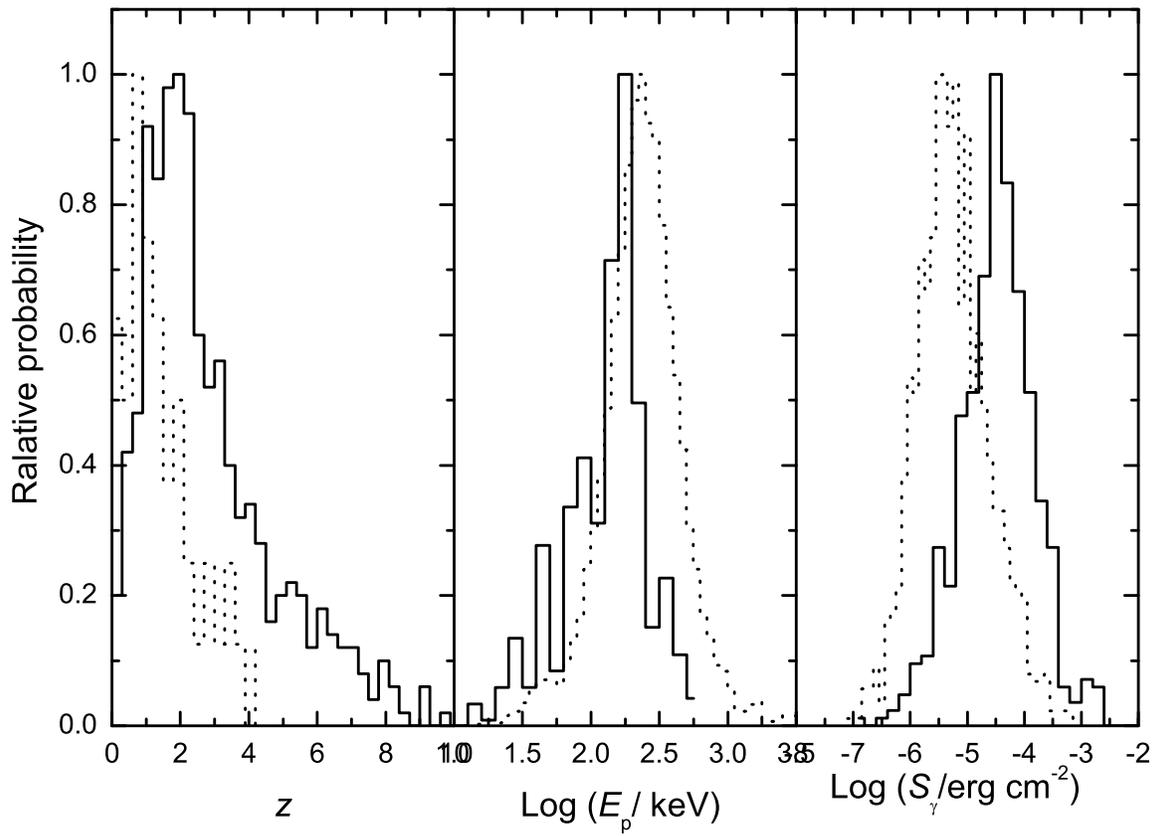} \caption{The distributions of $z$, $E_p$, and $S_{\gamma}$ for the
simulated GRB sample satisfying our model-independent standard candle relationship. For
comparison, the imposed dotted lines are the distributions derived from the observational
data. \label{fig8}}
\end{figure}

\begin{figure}
\plotone{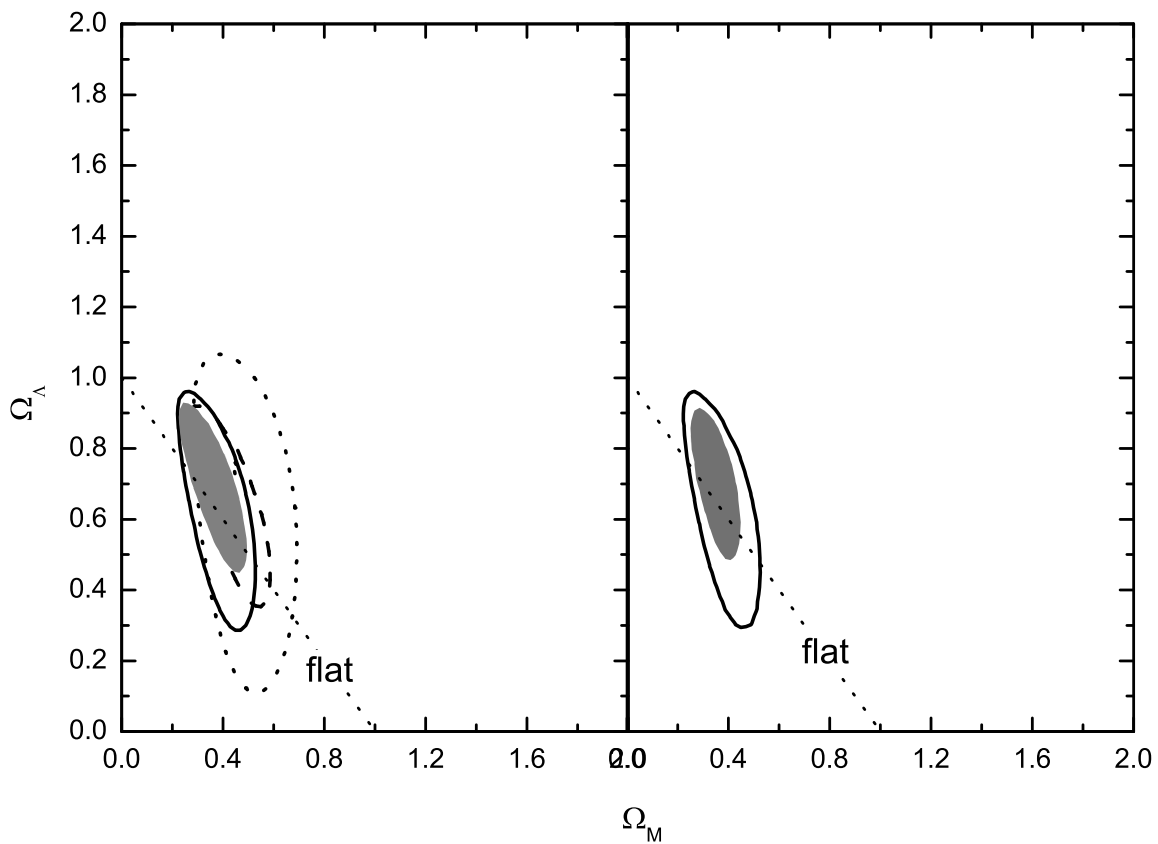} \caption{A comparison of the $1\sigma$ likelihood
contours for different simulated samples: {\em right---}  Simulations
for different sample sizes: dotted contour --- 25 GRBs,
solid contour --- 50 GRBs, dashed contour --- 75 GRBs, grey contour
--- 100 GRBs. The same observational errors ($\sigma_x/x=0.25k$, k is
a random number between $0\sim 1$) are adopted; {\em left:} Simulations for a same sample
size (50 GRBs) but for different observational errors: the line contour ---
$\sigma_x/x=0.25k$ and the grey contour--- $\sigma_x/x=0.15k$ (see the procedure of our
simulations). The dotted line is for a flat Universe. \label{fig9}}
\end{figure}

\begin{figure}
\plotone{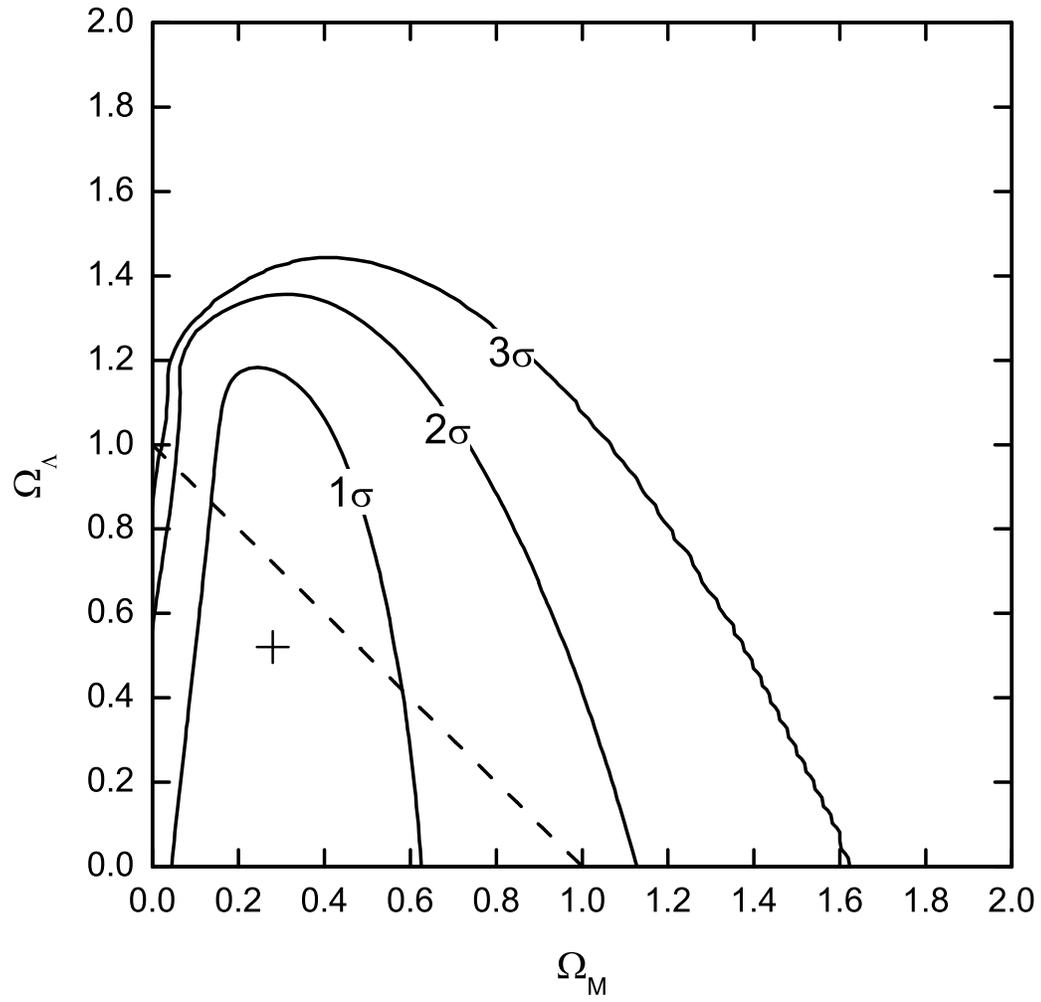} \caption{The contours of likelihood interval distributions in the
($\Omega_{\rm{M}}$, $\Omega_\Lambda$)-plane derived by the marginalization method. The
cross marks the most possible value of $(\Omega_{\rm{M}},\Omega_{\Lambda})$ , which is
$(0.28, 0.52)$. The contours give $0.05<\Omega_{\rm{M}}<0.61$($1\sigma$). Considering a
flat universe (the dashed line), the contours yield $0.14<\Omega_{\rm{M}}<0.58$ and
$0.40<\Omega_{\Lambda}<0.84$ ($1\sigma$). \label{fig10}}
\end{figure}

\begin{figure}
\plotone{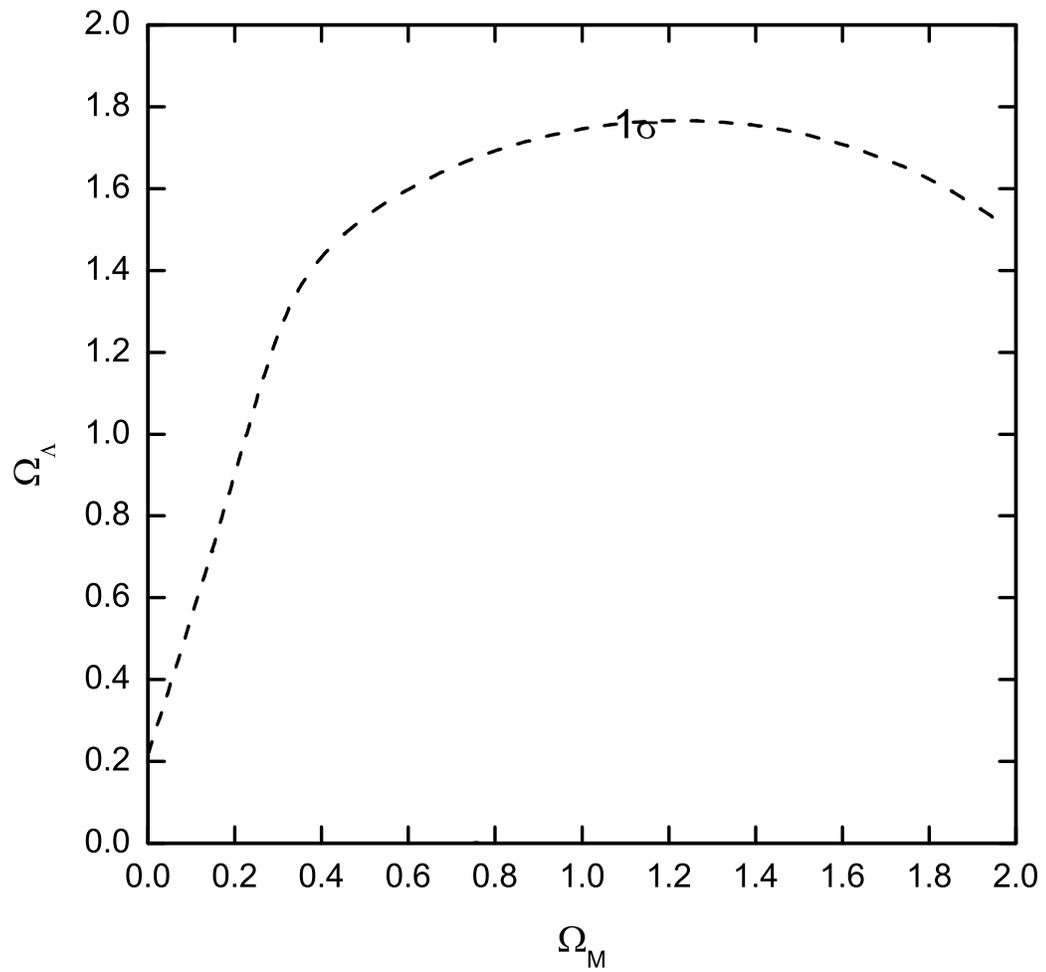} \caption{Same as Fig. \ref{fig5} but the uncertainties of the
parameters in the empirical relationships are also included into the error of the
distance modulus in the calculation of $p(\Omega)$ by Eq. \ref{PW}. Only $1\sigma$
interval is shown. \label{fig11}}
\end{figure}

Since launched on Nov. 20, 2004, the {\em Swift} mission (Gehrels et al. 2004) is
regularly detecting GRBs with a rate of $\sim 80$ bursts per year.
Detailed X-ray and UV/optical afterglow observations spanning from 1 minute to several
days after the burst have been performed for most of the bursts. However, the energy band
of the {\em Swift} Burst Alert Telescope (BAT) is narrow, i.e. 15-150 keV. As we shown in
Figure \ref{fig8}, the typical $E_p$ of a burst in our sample is marginally at the end of
the BAT energy band. As a result, BAT is not ideal for the purpose of expanding the
sample for GRB cosmology. As a result, we do not expect a dramatic enlargement of our
sample in the {\em Swift} era. Nonetheless, those bursts with an $E_p$ of $\sim (50-100)$
keV\footnote{Such a burst tend to be an X-ray rich GRB (Lamb et al. 2005a).} could have
their $E_p$ well-measured by {\em Swift}. We therefore highly recommend a detailed
optical follow-up observations for these bursts with UVOT and/or other ground-based
optical telescopes. This would present an opportunity to enlarge our sample with the {\em
Swift} data.

\begin{figure}
\plotone{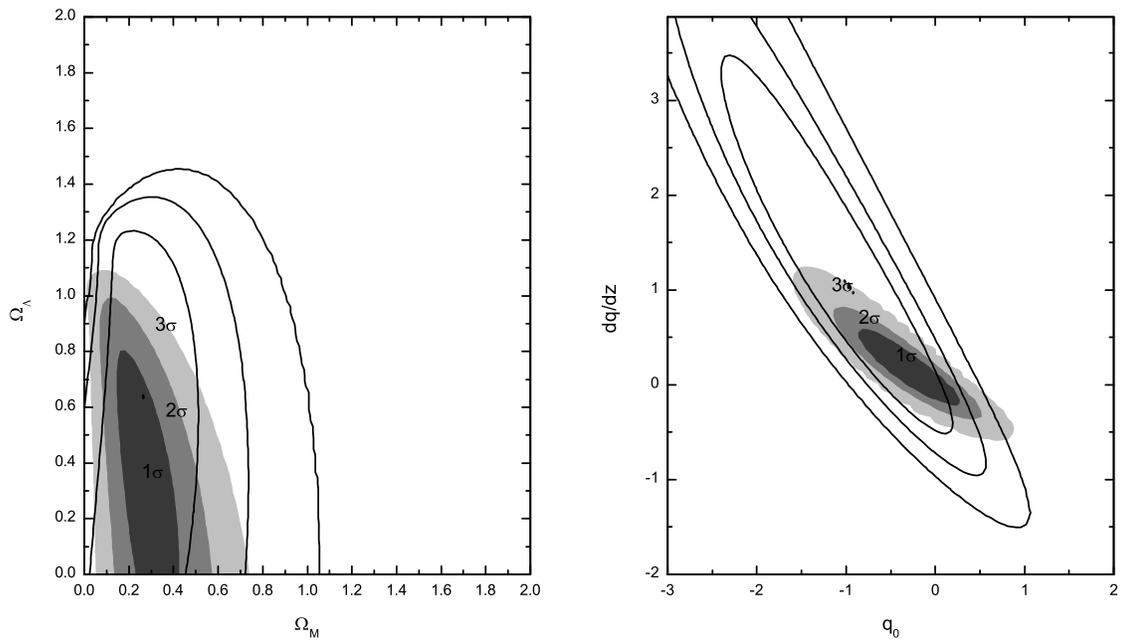} \caption{A demonstration of the potential constraints on cosmological
parameters with high redshift GRBs. The gray contours are the results derived from 5
pseudo high redshift GRBs together with the observed GRB sample [$1\sigma$ (dark gray) to
$3\sigma$ (light gray)], and the line contours are the results from the current observed
GRB sample only. \label{fig12}}
\end{figure}

The major advantage of GRBs serving as a standard candle over SNe Ia is their
high-redshift nature. The observed spectra and fluences of high redshift GRBs may not be
too different from the nearby ones (Lamb \& Reichart 2000). For example, the fluence of
GRB 000131 ($z=4.5$) is $~1\times 10^{-5}$ erg cm$^{-2}$ in the 25-100 keV band (Hurley
et al. 2000), which is significantly larger than the fluence of typical GRBs ($\sim
10^{-6}$ erg cm$^{-2}$ in the 25-2000 keV band). The highest redshift burst in our sample
is GRB 020124 ($z=3.2$).  Its fluence is $6.8 \times 10^{-6}$ erg cm$^{-2}$ in the 30-400
keV band, and its observed peak energy is 120 keV. These indicate that the current GRB
missions, such as {\em Swift} and {\em HETE-2}, are adequate to observe high redshift
GRBs\footnote{Strictly speaking, we refer to the optical band in the cosmological rest
frame to define $t^{'}_b$. This is not an issue if the GRB redshift is small. For
high-$z$ GRBs, the optical band in the observer's frame is highly extincted by neutral
hydrogen, but one could still detect $t_{\rm b}$ from the infrared band. Infrared band
observations are also essential for identify high-$z$ GRBs. IR follow up observations are
therefore essential to add the high-$z$ bursts in our sample.}. We explore how the
constraints on cosmological parameters are improved by identifying several high-redshift
bursts. We artificially select 5 high redshift GRBs from our simulated GRB sample with
$z\sim 4.0, \ 5.0, \ 6.0, \ 7.0 \ 8.0$, respectively, and with observational errors
$\sigma_x/x=0.25k$. The constraints on the cosmological parameters from these pseudo
high-z GRBs together with the current observed GRB sample are shown in Figure \ref{fig12}
(the grey contour), where the results from the current observed GRB sample are also
plotted for comparison (the line contours). It is found that adding a few high-$z$ GRBs
could result in much tighter constraints on cosmological parameters. Identifying high-$z$
GRBs and measuring their $E'_p$ and $t'_b$ are therefore essential for the GRB cosmology
in the near future.

Our model-independent relationship is close to the Ghirlanda
relationship, which was derived based on a simplest version of GRB jet
models. In such a model invoking a jet with energy uniformly
distributed in the jet cone, the observable $t_b$ is physically
related to the epoch when the bulk Lorentz factor of the ejecta is
reduced to the inverse of the jet opening angle (Rhoads 1999; Sari et
al. 1999). Relating $t_b$ to the jet openning angle, the jet energy is
then given by $E_{\rm{jet}} \propto (E_{\gamma,{\rm iso}} t_{\rm
b}^{'})^{3/4}(n\eta_\gamma)^{1/4}$, where $n$ is
circum-burst medium density and $\eta_{\gamma}$ is the efficiency of
GRBs. The Ghirlanda-relation can be then expressed as
$E_{\rm iso}\propto E_{\rm p}^{'2} t_{\rm
b}^{'-1}(n\eta_\gamma)^{-1/3}$. Comparing this with our
model-independent relationship (eq.\ref{MR}), we can see that both
relations are roughly consistent with each other if $n$ and
$\eta_{\gamma}$ are universal among bursts. As discussed above, the
motivations for us to introduce our multi-variable relationship are
two folds. Firstly, $n$ and $\eta_\gamma$ may not be constant and
actually vary from burst to burst. This introduces a lot more
uncertainties in the Ghirlanda relationship (e.g. Friedman \& Bloom
2005). Secondly and more importantly, there is no straightforward
interpretation of the relation within the jet model. Jumping out from
the jet model framework would give more freedom of theoretical
interpretations.

The tight relation of $E_{iso}(E_p^{'},t_b^{'})$ is very intriguing,
and its physical reason calls for investigation.
The fact that $E_{\gamma,{\rm{iso}}}$ strongly depends on
$E_{\rm {p}}^{'}$ and $t_{\rm{b}}^{'}$, while both $E_{{\rm {XA}},
{\rm{iso}}}$ and $E_{{\rm {OA}}, {\rm{iso}}}$ do not, implies
that $t_{\rm{b}}^{'}$ is a quantity related to GRBs rather
than to their afterglows. A similar signature was previously found by
Salmonson \& Galama (2002), who discovered a tight correlation between
the pulse spectral lag of GRB light curves and $t_b$. We therefore
suspect that $t_b$ might be a unique probe for the GRB prompt emission
properties. Within the jet scenario, the anti-correlation
between $t_b$ and $E_{\rm \gamma,iso}$ (first revealed by Frail et
al. 2001) may be physically related to the different metalicity
abundances of the progenitor stars (e.g. metal-poor stars rotate more
rapidly, and the GRBs they produce are more energetic and have more
collimated jets, MacFadyen \& Woosley 1999; Ramirez-Ruiz et al. 2002)
or may be simply a manifestation of the viewing angle effect in a
structured-jet scenario (Rossi et al. 2002; Zhang \& M\'esz\'aros
2002b). Such an anti-correlation, when combined with the physical
models of $E_{\rm \gamma,iso}-E_p$ correlations (e.g. Zhang \&
M\'esz\'aros 2002a; Rees \& M\'esz\'aros 2005), may be able to
interpret the observed $E_{iso}(E_p^{'},t_b^{'})$ relation, although
a detailed model is yet constructed. Alternatively, there might be
a completely different physical reason under the
$E_{iso}(E_p^{'},t_b^{'})$ relationship which is not attached to the
jet picture. One possibility is that the spectral break in the prompt
emission and the temporal break in the optical band may be related to
a same evolving break in the electron spectral distribution (B. Zhang,
2005, in preparation). In such an interpretation, the temporal break
time in the optical band is expected to be different from those in the
radio or in the X-ray bands. Since so far there is no solid proof for
the achromatic nature in broad bands for any ``jet break'',
such a possibility is not ruled out.

We are grateful to the anonymous referee for valuable
comments. We also thank Z.  G. Dai, D. Xu, Y. Z. Fan, X. F. Wu,
G. Rhee, D. Lamb, G. Ghirlanda, D. Lazzati, T. Piran, Y. P. Qin, and
B. B. Zhang for helpful discussions. This work is supported by NASA
NNG04GD51G, a NASA Swift GI (Cycle 1) program, and the National
Natural Science Foundation of China (No. 10463001).

\newpage

\begin{table*}
 \centering

 \caption{Prompt emission parameters of the GRB sample adopted in this
paper}
\begin{tabular}{llllllll}
\hline\hline%

GRB&$z$&$E_{\rm {p}}(\sigma_{E_{\rm
{p}}})$&$\alpha$&$\beta$&$S_{\gamma}(\sigma_S)$&Band &Refs.\\
(1)&(2)&(3)(keV)&(4)&(5)&(6)(erg.cm$^{-2}$)&(7)(keV)&(8)\\

\hline

980703&0.966&254(50.8)&-1.31&-2.40&22.6(2.3)&20-2000&1; 2; 2; 2\\
990123&1.6&780.8(61.9)&-0.89&-2.45&300(40)&40-700&3; 4 ; 4; 4\\
990510&1.62&161.5(16.1)&-1.23&-2.70&19(2)&40-700&5; 4; 4; 4\\
990712&0.43&65(11)&-1.88&-2.48&6.5(0.3)&40-700&5; 4; 4; 4\\
991216&1.02&317.3(63.4)&-1.23&-2.18&194(19)&20-2000&6; 2; 2; 2 \\
011211&2.14&59.2(7.6)&-0.84&-2.30&5.0(0.5)&40-700&7; 8; 8; 7\\
020124&3.2&86.9(15.0)&-0.79&-2.30&8.1(0.8)&2-400&9; 10; 10; 10\\
020405&0.69&192.5(53.8)&0.00&-1.87&74.0(0.7)&15-2000&11; 11; 11; 11\\
020813&1.25&142(13)&-0.94&-1.57&97.9(10)&2-400&12; 10; 10; 10\\
021004&2.332&79.8(30)&-1.01&-2.30&2.6(0.6)&2-400&13; 10; 10; 10\\
021211&1.006&46.8(5.5)&-0.86&-2.18&3.5(0.1)&2-400&14; 10; 10; 10\\
030226&1.986&97(20)&-0.89&-2.30&5.61(0.65)&2-400&15; 10; 10; 10\\
030328&1.52&126.3(13.5)&-1.14&-2.09&37.0(1.4)&2-400&16; 10; 10; 10\\
030329&0.1685&67.9(2.2)&-1.26&-2.28&163(10)&2-400& 17; 10; 10; 10\\
030429&2.6564&35(9)&-1.12&-2.30&0.85(0.14)&2-400&18; 10; 10; 10\\
\hline
\end{tabular}

\tablerefs{References are in order for $z$, $E_{\rm p}^{obs}$,
$[\alpha, \beta]$, $S_{\gamma}$:(1) Djorgovski et al.
1998; (2) Jimenez et al. 2001; (3) Kulkarni et al. 1999; (4) Amati et
al. 2002; (5) Vreeswijk et al. 2001;
(6)Djorgovski et al. 1999; (7) Holland et al. 2002; (8) Amati 2004;
(9) Hjorth et al. 2003; (10)Sakamoto et al. 2004b;
(11) Price et al. 2003; (12) Barth et al. 2003; (13) M\"oller et
al. 2002;  (14) Vreeswijk et al. 2003; (15) Greiner et
al. 2003; (16) Martini et al.  2003; (17)Bloom et al. 2003c; (18)
Weidinger et al. 2003.}

\end{table*}

\begin{table*}
 \begin{center}

 \caption{X-ray and optical afterglow data of the GRB sample adopted
in this paper}
\begin{tabular}{llllllllll}
\hline\hline%

GRB&$\alpha_x$\tablenotemark{a}&Epoch\tablenotemark{a}&$F_x$\tablenotemark{a}&$F_{x,10h}$($\sigma_{F_{x,10h}}$)\tablenotemark{a}&
$t_{\rm{b}}$($\sigma_{t_{\rm{b}}})$\tablenotemark{b}&$\alpha_1$\tablenotemark{b}&$\alpha_2$\tablenotemark{b}&Ref.\tablenotemark{b}&$R_{11h}$\tablenotemark{c}\\

(1)&(2)&(3)(hours) &(4)&(5)&(6)(days)&(7)&(8)&(9)&(10)\\

\hline

980703&1.24&34&4&18.24(4.97)&3.4(0.5)&-&-&1&20.1\\

990123&1.08&6&110&66.09(6.33)&2.04(0.46)&1.17&1.57& 2 & 19.4\\

990510&1.41&8.7&47.8&41.07(3.68)&1.6(0.2)&0.46&1.85&3&18.1\\

990712&-&-&-&-&1.6(0.2)&0.83&3.06&4&19.5\\

991216&1.61&4&1240&287.21(14.73)&1.2(0.4)&1&1.8& 5&16.9\\

011211&1.5&11&1.9&2.23(0.39)&1.56(0.02)&0.95&2.11&6&20.1\\

020124&-&-&-&-&3(0.4)&-&-&7&21.6\\

020405&1.15&41&13.6&68.98(20.21)&1.67(0.52)&1.4&1.95&8&18.3\\

020813&1.42&39&22&113.98(17.01)&0.43(0.06)&0.76&1.46& 9&19.1\\

021004&1.56&20.81&4.3&13.5(2.47)&4.74(0.14)&0.85&1.43&10&18.4\\

021211&-&-&-&-&1.4(0.5)&-&-&11&21.3\\

030226\tablenotemark{\$}&-&37.1&0.32&12.3&1.04(0.12)&0.77&1.99&12&19.5\\

030328\tablenotemark{\#}&-&15.33&3&-&0.8(0.1)&1.0&1.6& 13&20.2\\

030329\tablenotemark{@}&1.74&4.85&1400&467(23)&0.5(0.1)&1.18&1.81&14&14.7\\

030429&-&-&-&&1.77(1)&0.88&2.87&15&20.2\\

\hline

\end{tabular}
\end{center}
\tablenotetext{a}{Temporal decay index and X-ray afterglow flux in
2-10 keV band at a given observed epoch. $F_{x, 10h}$ is the
extrapolated/interpolated X-ray afterglow flux at 10 hours after the
GRB trigger. The fluxes are in units of $10^{-13}$ ergs cm$^{-2}$ s
$^{-1}$. They are taken from Berger et al. (2003) except for those with
marks: 030226 (Pedersenet al. 2003); 030328 (Butler et al. 2003);
030329 (Marshall \& Swank 2003; Marshall, Markwardt, \& Swank 2003;
Tiengo, Mereghetti, \& SchartelA 2003a, b)}

\tablenotetext{b}{Temporal break (error) and temporal indices before
and after the break, and their references: (1) Frail et
al. 2003; (2) Kulkarni et al. 1999; (3) Stanek et al. 1999;
(4) Bj\"{o}rnsson et al. 2001; (5) Halpern et al. 2000; (6) Jakobsson et
al. 2003; (7) Berger et al. 2002; (8) Price et al. 2003; (9) Barth et
al. 2003; (10) Holland et al. 2003; (11) Holland et al. 2004; (12) Klose
et al. 2004; (13) Andersen et al. 2003; (14) Berger et al. 2003;
(15) Jakobsson et al.  2004a.}

\tablenotetext{c}{R-band magnitude adjusted to 11 hours after the burst
trigger (from Jakobsson et al. 2004b).}

\end{table*}

\begin{table*}
 \centering

 \caption{Derived isotropic energies, rest frame peak energies and
rest frame temporal breaks for the GRB sample adopted in this paper
(Assuming $\Omega_{\rm{M}}=0.28$ and $\Omega_\Lambda=0.72$)}
\begin{tabular}{llllll}
\hline\hline%

GRB&$\log E_{\gamma,{\rm{iso}}}(\sigma_{E_{\gamma}})$&$\log E_{\rm
XA,{\rm{iso}}}$&$\log E_{\rm OA,{\rm{iso}}}$&$\log E^{'}_{\rm
p}(\sigma_{E_{\rm {p}}})$&$\log t^{'}_{b}(\sigma_{t_{\rm{b}}})$\\
&(1)(erg)&(2)(erg)&(3)(erg)&(4)(keV)&(5)(day)\\

\hline

980703&52.85(0.04)&47.69&-&2.70(0.09)&0.238(0.064)\\

990123&54.64(0.06)&48.80&46.09&3.31(0.03)&-0.105(0.098)\\

990510&53.29(0.05)&48.58&46.60&2.63(0.04)&-0.222(0.008)\\

990712&51.88(0.02)&-&44.08&1.97(0.07)&0.049(0.054)\\

991216&53.85(0.04)&49.00&45.70&2.81(0.09)&-0.226(0.145)\\

011211&53.01(0.04)&47.72&45.84&2.27(0.06)&-0.304(0.006)\\

020124&53.37(0.05)&-&-&2.70(0.08)&-0.146(0.058)\\

020405&53.17(0.01)&47.92&45.91&2.51(0.12)&-0.005(0.135)\\

020813&54.13(0.06)&48.90&45.42&2.68(0.09)&-0.719(0.061)\\

021004&52.66(0.10)&48.54&46.82&2.42(0.16)&0.153(0.013)\\

021211&52.05(0.03)&-&-&1.97(0.05)&-0.156(0.155)\\

030226&52.90(0.05)&47.52&46.14&2.46(0.09)&-0.458(0.050)\\

030328&53.60(0.02)&47.66&45.51&2.50(0.05)&-0.498(0.054)\\

030329&52.19(0.04)&48.27&45.40&1.90(0.01)&-0.369(0.087)\\

030429&52.24(0.07)&-&46.30&2.11(0.11)&-0.315(0.245)\\

\hline

\end{tabular}

\end{table*}

\begin{table*}\label{Table4}
\centering

\caption{The results of multiple variable regression analysis
(Assuming $\Omega_{\rm{M}}=0.28$ and $\Omega_\Lambda=0.72$)}
\begin{tabular}{cccc}

\hline\hline
  &$\hat{E}_{\gamma,\rm{{\rm{iso}}}}(E_{p}^{'},t_{b}^{'})$&$\hat{E}_{\rm{X,{\rm{iso}}}}(E_{p}^{'},t_{b}^{'})$&$\hat{E}_{\rm{R,{\rm{iso}}}} (E_{p}^{'},t_{b}^{'})$\\
\hline
$\kappa_0(p_t)$& $48.0\pm0.4(<10^{-4})$  &$46.27\pm 1.35(<10^{-4})$  &$44.22\pm 1.42(<10^{-4})$\\
$\kappa_1(p_t)$& $1.94\pm0.17(<10^{-4})$  & $0.74\pm 0.51(0.18)$ &$0.64\pm 0.57(0.26)$\\
$\kappa_2(p_t)$& $-1.24\pm0.23(2\times 10^{-4})$  & $-0.30\pm 0.62 (0.64)$ &$0.29\pm 0.88 (0.75)$\\
\hline
Global F-test statistics & 115.4   & 1.08   &0.77\\
and probability $p_F$&$<10^{-4}$ & 0.39& 0.49\\
\hline
Global correlation $r$ & $0.96\pm 0.21$&  $0.46\pm 0.24$ &  $0.38 \pm 0.28$  \\
and probability $P_S$& $<10^{-4}$ &0.15 & 0.22\\
\hline
\end{tabular}

\end{table*}

\newpage

\end{document}